\documentclass[useAMs,usenatbib,a4paper]{mn2e}
\usepackage{savesym}
\usepackage{graphicx}
\expandafter\let\csname equation*\endcsname\relax
  \expandafter\let\csname endequation*\endcsname\relax 
\usepackage{subfig}
\usepackage{amsmath}
\usepackage{amssymb}
\usepackage{verbatim}
\usepackage{array}
\usepackage{times}
\usepackage[total={17.8cm,24.0cm},centering]{geometry} 

\newcommand{\be}{\begin{equation}}
\newcommand{\ee}{\end{equation}}
\newcommand{\eea}{\end{eqnarray}}
\newcommand{\bea}{\begin{eqnarray}}

\newcommand{\m}{\mathrm}

\newcommand{\grays}{$\gamma$-rays}

\title[Compton-dominant blazars]{Synchrotron and inverse-Compton emission from blazar jets - III. Compton-dominant blazars}
\author[William J. Potter and Garret Cotter]{William J. Potter\thanks{E-mail:
will.potter@astro.ox.ac.uk (WJP)} and Garret Cotter
\\
Oxford Astrophysics. Denys Wilkinson Building, Keble Road, Oxford, OX1 3RH, United Kingdom}
\begin{document}

\date{}

\pagerange{\pageref{firstpage}--\pageref{lastpage}} \pubyear{2012}

\maketitle

\label{firstpage}

\begin{abstract}
In this paper we develop the extended jet model of Potter $\&$ Cotter to model the simultaneous multi-wavelength spectra of six Compton-dominant blazars.  We include an accelerating parabolic base transitioning to a slowly decelerating conical jet with a geometry set by observations of M87 and consistent with simulations and theory.  We investigate several jet models and find that the optically thick to thin synchrotron break in the radio spectrum requires the jet to first come into equipartition at large distances along the jet, consistent with the observed transition from parabolic to conical at $10^{5}R_{s}$ in the jet of M87.  We confirm this result analytically and calculate the expected frequency core-shift relations for the models under consideration.  We find that a parabolic jet transitioning to a ballistic conical jet at $10^{5}R_{s}$, which starts in equipartition and becomes more particle dominated at larger distances, fits the multiwavelength data of the six blazars well, whilst an adiabatic equipartition conical section requires very large bulk Lorentz factors to reproduce the Compton-dominance of the blazars.

We find that all these blazars require high power ($>10^{39}W$), high bulk Lorentz factor ($>20$) jets observed close to the line of sight ($<2^{o}$) as we expect from the blazar sequence and consistent with the results from Paper II.  The inverse-Compton emission in our fits is due to inverse-Compton scattering of high-redshift CMB photons at large distances along the jet due to the high bulk Lorentz factors of the jets.  We postulate a new interpretation of the blazar sequence based on the radius of the transition region of the jet (where the jet is brightest in synchrotron emission) scaling linearly with black hole mass.

\end{abstract}


\section{Introduction}

Blazars are the most luminous active galactic nuclei (AGN).  Superluminal motion is observed in their jets indicating that the jets are relativistic and observed at small angles to the line of sight.  The emission from the relativistic jets is strongly Doppler-boosted so that the non-thermal radiation overwhelms the observed emission from the host galaxy.  High energy electrons in the plasma jets emit synchrotron and inverse-Compton radiation resulting in the characteristic double-peaked spectrum, with synchrotron radiation from radio to UV/x-rays and inverse-Compton emission from x-rays to very high energy \grays.  Blazars are observed to have self-absorbed synchrotron radio spectra which are nearly flat in flux.

In the first two papers in this series  (\cite{2012MNRAS.423..756P} and \cite{2013MNRAS.429.1189P}, hereafter Paper I and Paper II respectively) we developed an extended jet model based on recent VLBI observations of M87, numerical simulations and theory.  In Paper I we introduced a simple ballistic conical jet model which we showed could successfully reproduce the quiescent spectrum of BL Lacertae across all wavelengths using a relatively low power, low bulk Lorentz factor jet.  We included a thorough treatment of emission processes by integrating the line of sight synchrotron optical depth through the jet and numerically integrating the exact expression for synchrotron self-Compton (SSC) emission. In Paper II we developed this model to include an accelerating parabolic base transitioning to a slowly decelerating conical jet motivated by radio observations of M87 (\cite{2011arXiv1110.1793A}).  We included inverse-Compton emission from synchrotron, CMB, accretion disc, starlight, broad line region (BLR), dusty torus and narrow line region (NLR) scattered seed photons by Lorentz transforming the photon distribution into the plasma rest frame and using the full Klein-Nishina cross-section.   

We used this model to successfully fit to the quiescent spectrum of the most Compton-dominant object in the {\sl FERMI} sample PKS0227-369 (\cite{2010ApJ...716...30A}), using a jet geometry fixed by the radio observations of M87 (\cite{2011arXiv1110.1793A}) scaled linearly with black hole mass.  We found that the fit to PKS0227 required a high power, high bulk Lorentz factor jet observed close to the line of sight, consistent with our expectations from the blazar sequence (\cite{1998MNRAS.299..433F}) and unification (\cite{1995PASP..107..803U}).  We found that the spectrum of PKS0227 was well fitted by a jet which comes into equipartition outside of the BLR; within the dusty torus or at a larger distance along the jet.  We found that the inferred black hole mass of our fit was close to a previous estimate using a transition region (where the jet comes into equipartition and the jet transitions from parabolic to conical) at a distance of $10^{5}R_{s}$, as observed in the jet of M87.  For this fit the transition region occured at a distance 34pc from the central black hole where the Compton-dominance of the blazar was due to inverse-Compton scattering of high-redshift CMB photons which were strongly Doppler-boosted into the plasma rest frame due to the jet\rq{}s high bulk Lorentz factor.  We postulated that one of the reasons Compton-dominant blazars tend to be observed at large redshifts is due to the temperature dependence of the CMB on redshift ($T_{\m{CMB}}\propto (1+z)$).

In this paper we set out to investigate whether this model is capable of fitting to the spectra of a sample of Compton-dominant blazars and to investigate what physical properties these blazars have in common.  We wish to find out whether the physical parameters we found for PKS0227 in Paper II hold for other Compton-dominant blazars, in particular, if we can distinguish between models in which the transition region occurs within the dusty torus or further along the jet using the optically thick to thin synchrotron break.  This is the first time a realistic extended jet model has been used to fit to the spectra of a sample of blazars, so it is interesting to see whether our model, which is based on the observed geometry of the jet in M87, is capable of fitting to a sample of real blazars.  For this sample we choose the six most Compton-dominant blazars (after PKS0227) from \cite{2010ApJ...716...30A} in order to test our model and predictions from Paper II.

Compton-dominant blazars are thought to represent the most intrinsically powerful jets in the blazar sequence (\cite{1998MNRAS.299..433F}). In this scenario we use the radio luminosity of blazars as a proxy for their kinetic luminosity and take an average of the multi-wavelength spectra in bins of radio power.  A general progression towards higher Compton-dominance and decreasing peak frequency of synchrotron and inverse-Compton emission  for more powerful objects was found by \cite{1998MNRAS.299..433F}.  It is currently believed that the Compton-dominance of powerful blazars originates from inverse-Compton scattering of external photons (\cite{2012arXiv1203.4991M}).  It is thought that powerful objects with more luminous accretion discs and broad line regions (BLRs) create a larger seed photon distribution for inverse-Compton scattering than in BL Lac type objects where the dominant photon source is synchrotron seed photons.  It is believed that radiative cooling of accelerating electrons via inverse-Compton scattering of BLR photons leads to the lower inverse-Compton peak frequencies in Compton-dominant objects (\cite{2008MNRAS.391.1981M}).  

In Paper II we considered scattering of synchrotron, accretion disc, BLR, dusty torus, NLR, starlight and CMB photons and found that PKS0227 was best fitted by inverse-Compton scattering of CMB or dusty torus seed photons.  We found that the low inverse-Compton peak frequency was due to the lower peak energy of CMB and dusty torus photons relative to synchrotron seed photons.  We found that the Compton-dominance was due to the increase in external photon density due to Doppler-boosting in the high bulk Lorentz factor jet.  The low synchrotron peak frequency was due to the base of the conical section of the jet having a lower magnetic field due to the high black hole mass and the assumed linear scaling of the M87 jet geometry with black hole mass.  This was the first time that such a scenario has been suggested for the Compton-dominance of blazars, however, this was also the first time that a realistic model of an extended jet had been used to model the spectrum of a Compton-dominant blazar.   

In this paper we will first briefly introduce our jet model.  We compare several adiabatic and ballistic jet models to the radio emission of the six most Compton-dominant blazars in \cite{2010ApJ...716...30A} (after PKS0227).  We use the slope of the radio emission and the frequency of the optically thick to thin synchrotron break to distinguish between a transition region within the dusty torus or at futher distances along the jet.  We then show the results of fitting the surviving models to the multi-wavelength spectra of the six blazars.  We compare the physical parameters inferred from the model fits to the results of Paper II in order to determine whether our model is capable of reproducing the quiescent spectra of a sample of blazars.  We comment on our results in light of current ideas on the blazar population and unification.

\section{Jet Model}

To fit to the spectra of the Compton-dominant blazars we use a model with an accelerating parabolic base transitioning to a slowly decelerating conical jet with a geometry set by the recent radio observations of the jet of M87 (\cite{2011arXiv1110.1793A}).  We will briefly describe the principles of this model in the following section (for full details see Papers I and II).  

Our model is motivated by and consistent with observations, simulations and theory.  The jet is assumed to start magnetically dominated at the base.  The magnetic energy is converted into bulk kinetic energy in the plasma via a magnetic pressure gradient.  The plasma is accelerated until it approaches equipartition between magnetic and particle energies where the jet approaches a terminal bulk Lorentz factor and transitions from parabolic to conical (\cite{2006MNRAS.368.1561M}).  In the conical section the jet slowly decelerates due to interactions with its environment (\cite{2002MNRAS.336..328L}, \cite{2005MNRAS.358..843H} and \cite{2010ApJ...710..743D}) and this converts bulk kinetic energy into accelerating electrons via shocks.  Deceleration is observed along jets and this is also consistent with evidence for in situ acceleration of electrons from observations of optical synchrotron emission (\cite{1997A&A...325...57M} and \cite{2001A&A...373..447J}).  We assume that the plasma is isotropic and homogeneous in the rest frame of each section of plasma so can be described by a relativistic perfect fluid.  We conserve energy-momentum and electron number via the conservation equations

\be
\nabla_{\mu}T^{\mu \nu}(x)=0, \qquad \nabla_{\mu}j_{e}^{\mu}(x)=0.
\ee

We calculate the synchrotron emission and opacity in each section by Lorentz transforming into the instantaneous plasma rest frame.  We integrate the synchrotron optical depth through the jet to each section to calculate the observed synchrotron emission.  In the model we treat inverse-Compton scattering of synchrotron, CMB, starlight, accretion disc, BLR, dusty torus and NLR seed photons by Lorentz transforming into the plasma rest frame and using the full Klein-Nishina cross-section.  We evolve the electron population along the jet taking into account radiative losses from synchrotron and inverse-Compton emission and adiabatic losses due to expansion of the jet.  We set the geometry of our model using the observations of the geometry of M87 with lengths scaled linearly with black hole mass.  
 
\subsection{Model parameters}

\begin{figure}
	\centering
		 \includegraphics[width=7 cm, clip=true, trim=4cm 2cm 8cm 4cm]{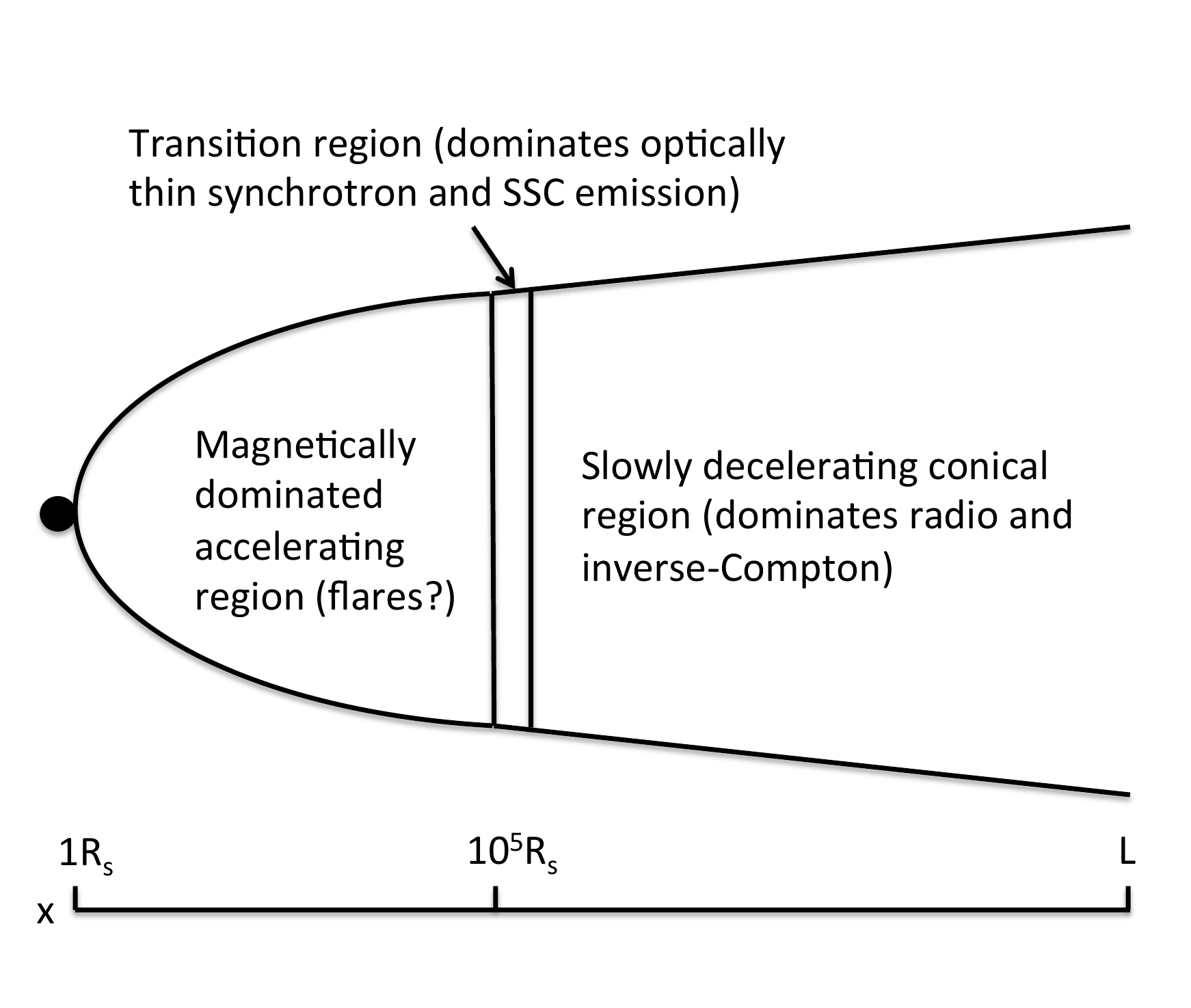} 
			
	\caption{This figure shows a schematic diagram of our jet model. }
	\label{fig:jet}
\end{figure}

We consider a jet with total lab frame power $W_{j}$, jet length $L$ and conical half-opening angle $\theta\rq{}_{\m{opening}}$ observed at an angle $\theta_{\m{observe}}$ to the jet axis.  We assume that electrons accelerated due to shocks (or via magnetic reconnection) along the jet have an initial distribution, $N_{e}(E_{e})\propto E_{e}^{-\alpha} \m{e}^{-E_{e}/E_{\m{max}}}$, with a minimum energy $E_{\m{min}}$.  The base of the jet has an initial bulk Lorentz factor $\gamma_{0}$, which accelerates to $\gamma_{\m{max}}$ at the end of the parabolic section.  The jet slowly decelerates from a bulk Lorentz factor of $\gamma_{\m{max}}$ to $\gamma_{\m{min}}$ by the end of the conical section.  The central black hole has a mass $M$.  A schematic diagram of the jet is shown in Figure $\ref{fig:jet}$.

\subsection{Electron energy losses due to emission}

In this paper we have modified our treatment of electron energy losses from Paper II to include the decrease of individual electron energies with radiation losses. In Paper II we included radiative energy losses to the electron population due to both synchrotron and inverse-Compton emission, however, we neglected the decrease of electron energy due to radiative losses and used the approximation that electrons radiate the majority of their emission close to their initial electron energy. This approximation resulted in an increase in speed of our jet code and did not produce a significant change in the emitted spectrum of the jet, however, the value of $E_{\m{max}}$ is underestimated for a fit to a spectrum. To include the transfer of electrons down through energy bins as they radiate we modify Equation 59 from Paper II to become

\bea
N_{e}(E_{e},x+\gamma_{\m{bulk}}\m{d}x')=N_{e}(E_{e},x)-\frac{P_{tot}(x,\m{d}x',E_{e})\times 1\m{m}}{cE_{e}(1-1/\m{d}E_{e})}\nonumber \\+\frac{P_{tot}(x,\m{d}x',E_{e}\m{d}E_{e})\times 1\m{m}}{c E_{e}\m{d}E_{e}(1-1/\m{d}E_{e})}.
\label{Etrans}
\eea

where $N_{e}(E_{e},x+\gamma_{\m{bulk}}\m{d}x')$ is the number of electrons of energy $E_{e}$ contained in a slab of width 1m (in the x-direction) in the plasma rest frame, $P_{tot}(x,\m{d}x',E_{e})$ is the total power radiated by electrons of energy $E_{e}$ through synchrotron and inverse-Compton emission in the jet section located at a lab frame distance $x$ with a width $\m{d}x$' in the plasma rest frame and $dE_{e}$ is the fractional difference in energies between adjacent electron energy bins in our code. We have also corrected the typo in Equations 10 and 59 of Paper II and changed 1s to 1m. The first term in Equation \ref{Etrans} corresponds to the number of electrons lost from the energy bin through radiative energy losses when crossing the section of width $\m{d}x'$ in the plasma rest frame. The second term corresponds to the increase in the number of electrons in the energy bin due to higher energy electrons cascading down in energy following radiative losses. If choosing to take into account the decrease in electron energy as the electron radiates, Equation 10 in Paper II should also be modified to include the electron gain and loss terms above.

\section{Results}

In this section we shall try to constrain the properties of the jet by fitting our model by eye to the simultaneous multi-wavelength observations of the six most Compton-dominant blazars (after PKS0227) J0349.8$-2102$, J0457.1$-2325$, J0531.0$+1331$, J0730.4$-1142$, J1504.4$+1030$ and J1522.2$+3143$ from \cite{2010ApJ...716...30A}.  First we compare the radio observations with the calculated radio emission from several different jet models.  We then use the optically thick to thin synchrotron break to attempt to constrain the location of the transition region of the jet.  We calculate an analytic formula for the observed synchrotron break frequency of a general jet plasma described by a relativistic perfect fluid.  We use this to confirm the results of our model and to infer the properties of parabolic and conical jets by comparing our calculations with observations of the frequency dependent core-shift in jets.  Finally, we fit the multiwavelength spectra using the jet models compatible with the radio observations and optically thick to thin synchrotron break.  Throughout this paper we assume a standard $\Lambda$CDM cosmology with $H_{0}=71 \m{km}\m{s}^{-1}\m{Mpc}^{-1}$ and $\Omega_{\Lambda}=0.73$. 

\subsection{The different models}

In this Paper we wish to constrain the physical properties of the jet by fitting different models to the observations of the six blazars.  In Paper II we found that a ballistic equipartition jet best fitted the observations of PKS0227, whilst an adiabatic equipartition jet struggled to reproduce its Compton-dominance with plausible bulk Lorentz factors.  Here we will investigate models in which the conical section of the jet is adiabatic or ballistic with a transition region located outside of the BLR.  In these models we will also consider cases where the plasma deviates from equipartition along the conical section, however, we still assume that the transition region is in equipartition.   

\begin{figure*}
	\centering
		\subfloat[Radio emission of J0457]{ \includegraphics[width=8 cm, clip=true, trim=1cm 1cm 0cm 1cm]{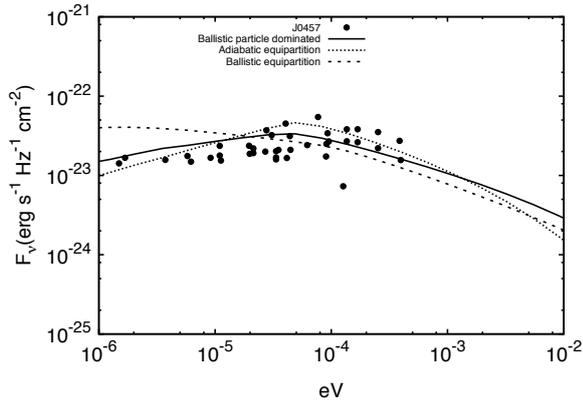} }
		\qquad
		\subfloat[Optically thick to thin synchrotron break in J0531]{ \includegraphics[width=8cm, clip=true, trim=1cm 1cm 0cm 1cm]{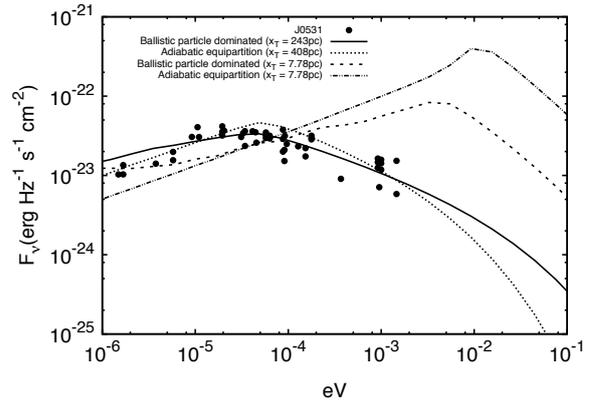} }
			
	\caption{Figure a shows the results of fitting different jet models to the radio emission of J0457.  We see that the ballistic equipartition jet does not match the radio observations well at low frequencies.  The adiabatic equipartition model fits the optically thick radio slope well as does a ballistic model which becomes particle dominated beyond the transition region ($U_{e}/U_{B} \propto x^{-1}$ from $x_{T}\leq x \geq 100x_{T}$).  In Figure b we show fits to the radio emission of J0531 using adiabatic and ballistic jet models with a transition region located at $x_{T}$ within the dusty torus and at larger distances along the jet (as implied by a transition region at $10^{5}R_{S}$ in M87).  The change in gradient of the flux from optically thick to optically thin at $3\times10^{-5}\m{eV}$ is well matched by transition regions at large distances along the jet $\sim100\m{pc}$ and excludes a transition region within the dusty torus at a few pc since the jet remains optically thick up to much higher frequencies than the observed break.   }
	\label{fig:1}
\end{figure*}

\section{Radio emission}

In Paper II we found that the spectrum of PKS0227 was best fitted by models in which the conical section of the jet was ballistic.  In Figure \ref{fig:1}a we show the radio emission from three models compared to the radio data of J0457.  We find that the radio observations of these six Compton-dominant blazars have radio slopes which are in general steeper than the radio emission produced by the ballistic equipartition jet model from Paper II, as shown in Figure \ref{fig:1}a.  We find that the ballistic equipartition jet tends to overproduce low frequency radio emission compared to the data.  This suggests that if the jet is ballistic and conical it becomes particle dominated outwards from the transition region since this reduces the radio synchrotron emission at larger distances, whilst maintaining the Compton-dominance from scattering external photons.  The radio emission from both the adiabatic jet in equipartition and the ballistic jet which becomes particle dominated fit the radio data well, whilst the ballistic equipartition model does not.  Radio observations are usually neglected in most previous investigations modelling blazar spectra.  These results, however, show that it is useful to consider the radio emission when constraining the physical properties of jets.  In Paper II we struggled to reproduce the observed Compton-dominance of PKS0227 using an adiabatic conical jet in equipartition without using very high bulk Lorentz factors. This suggests that the conical section of the jet may be ballistic starting in equipartition at the transition region and becoming more particle dominated further down the jet as magnetic energy continues to dissipate and accelerate electrons.

We find that in order to reproduce the observed radio slope of these blazars the equipartition fraction of the ballistic jet should decrease roughly as $A_{\m{equi}}=U_{B}/U_{e}=2x_{T}/(x_{T}+x)$ for $x_{T}\geq x\geq 100x_{T}$ (where $x_{T}$ is the distance of the transition region) as shown in Figure \ref{fig:1}a.  The radio data does not constrain the emission from the jet at distances beyond $\sim$10kpc and so we find that the ballistic jet becomes particle dominated by a factor of 50 at a distance of $100x_{T}$ but is not well constrained after this.  The ballistic jet could stay particle dominated, tend back towards equipartition or become even more particle dominated after this distance. 

\section{Optically thick to thin synchrotron break}

In Paper II we found that models in which the transition region either occured within the dusty torus or at further distances along the jet were able to fit well to the spectrum of PKS0227.  The main difference between the two models was the location of the optically thick to thin synchrotron break due to the different radii and magnetic field strengths of the transition regions in the two fits.  Figure \ref{fig:1}b shows the fits to the spectrum of J0531 in which the jet has a transition region located within the dusty torus and further along the jet corresponding to a distance $10^{5}R_{s}$ as in the jet of M87.  We see that in order to reproduce the relatively low frequency of the optically thick to thin synchrotron break at $\approx3\times10^{-5}\m{eV}$, requires a transition region outside the dusty torus at a distance of several 100pc.  If the jet comes into equipartition within the dusty torus then the jet overproduces the radio emission due to the optically thick to thin synchrotron break occuring at a higher frequency than indicated by the observations.  This result holds for both the ballistic jet which becomes particle dominated and the adiabatic equipartition jet.  This is interesting since for these large distances the inferred black hole mass of the jet agrees with the black hole mass found independently from fitting the accretion disc spectrum to J0531 only if the transition region occurs at $10^{5}R_{s}$, as we have assumed based on observations of M87.  It is worth noting that we find it difficult to reproduce both the radio and gamma-ray data using a jet model with a transition region within the dusty torus. 

The frequency at which the synchrotron emission becomes optically thin can be calculated analytically in the case of a jet with a total jet power $W_{j}$, bulk Lorentz factor $\gamma_{\m{bulk}}$, radius $R$, and equipartition fraction $A_{\m{equi}}=U_{B}/U_{e}$.  In this case the synchrotron opacity can be calculated using the formula from Paper I, assuming the plasma can be approximately described as a relativistic perfect fluid in the plasma rest frame (see Paper II) and the synchrotron emission is approximated to occur at the synchrotron critical frequency (see Paper I).

\be
k_{\nu}=\frac{j_{\nu}c^{2}}{2\epsilon^{1/2}\nu^{5/2}}, \qquad E_{e}^{2}=\epsilon \nu, \label{knu}
\ee

\be
j_{\nu}=\frac{\sigma_{T}AB^{2}\epsilon^{\frac{3-\alpha}{2}}\nu^{\frac{1-\alpha}{2}}}{3 \pi R^{2}\mu_{0}m_{e}^{2}c^{3}}, \qquad \epsilon=\frac{4 \pi m_{e}^{3}c^{4}}{3eB}. \label{jnu}
\ee

We can parameterise $A$, which is the normalisation constant for the electron distribution in a slab of width 1m ($N_{e}=AE^{-\alpha}$), in terms of the power law index of the electron distribution $\alpha$ and the maximum and minimum electron energies $E_{\m{min}}$ and $E_{\m{max}}$.  Similarly we can express the magnetic field strength in terms of other known quantities.

\be
B=\sqrt{\frac{3W_{j}\mu_{0}A_{\m{equi}}}{2\gamma_{\m{bulk}}^{2}c \pi R^{2}(1+A_{\m{equi}})}},
\ee

\be
A=\frac{3W_{j}}{4\gamma_{\m{bulk}}^{2}c(1+A_{\m{equi}})}.\left[\frac{(2-\alpha)}{E_{\m{max}}^{2-\alpha}-E_{\m{min}}^{2-\alpha}})\right].
\ee

In the case of $\alpha=2$ the factor $(2-\alpha)/(E_{\m{max}}^{2-\alpha}-E_{\m{min}}^{2-\alpha})$ should be replaced by $1/\ln (E_{\m{max}}/E_{\m{min}})$ in the above and following Equations.  We can now substitute these values into Equations \ref{knu} and \ref{jnu} to find an expression for the synchrotron opacity.

\bea
k_{\nu}=\frac{\sigma_{T}}{8 \pi \mu_{0}m_{e}^{2}c^{2}(1+A_{\m{equi}})}.\frac{2-\alpha}{E_{\m{max}}^{2-\alpha}-E_{\m{min}}^{2-\alpha}}.\left(\frac{4\pi m_{e}^{3}c^{4}}{3e}\right)^{\frac{2-\alpha}{2}} \nonumber \\
\nonumber \\
\times \left(\frac{3\mu_{0}A_{\m{equi}}}{2c\pi (1+A_{\m{equi}})}\right)^{\frac{2+\alpha}{4}}.\frac{W_{j}^{\frac{6+\alpha}{4}}\nu^{\frac{-(4+\alpha)}{2}}}{R^{\frac{6+\alpha}{2}}\gamma_{\m{bulk}}^{\frac{6+\alpha}{2}}}. \label{knu2}
\eea

This expression is quite general and is valid for any homogeneous plasma which is described as a relativistic perfect fluid in the plasma rest frame.  In our present discussion we are interested in calculating the jet radius at which a given synchrotron frequency becomes optically thick, for a plasma in equipartition. We assume that the plasma becomes optically thick to a frequency when $k_{\nu}d\m{x}=1$, where both $\nu$ and $d\m{x}$ are measured in the plasma rest frame.  The lifetime of electrons emitting synchrotron radiation at radio frequencies are long, so the length scale $d\m{x}$ will be determined such that the jet radius and magnetic field strength change significantly, reducing the synchrotron opacity.  We take this length scale to be the distance over which the jet radius doubles.  For a conical jet this requirement corresponds roughly to $d\m{x}=R$ because in the lab frame the conical half opening angle is expected to be approximately $\theta=1/\gamma_{\m{bulk}}$ and the lab frame length is related to that in the plasma rest frame by a length contraction $d\m{x}\rq{}=\gamma_{\m{bulk}}d\m{x}$.  Calculating $k_{\nu}R=1$ and rearranging we find

\bea
R=\left[\frac{\sigma_{T}}{8 \pi \mu_{0}m_{e}^{2}c^{2}(1+A_{\m{equi}})}.\frac{2-\alpha}{E_{\m{max}}^{2-\alpha}-E_{\m{min}}^{2-\alpha}}.\right]^{\frac{2}{4+\alpha}}. \nonumber \\
\left[\left(\frac{4\pi m_{e}^{3}c^{4}}{3e}\right)^{\frac{2-\alpha}{2}}\left(\frac{3\mu_{0}A_{\m{equi}}}{2c\pi (1+A_{\m{equi}})}\right)^{\frac{2+\alpha}{4}}\right]^{\frac{2}{4+\alpha}}.\frac{W_{j}^{\frac{6+\alpha}{2(4+\alpha)}}}{\gamma_{\m{bulk}}^{\frac{6+\alpha}{4+\alpha}}\nu}. \label{Rcon}
\eea

We see that for a conical jet in which the electron distribution, equipartition fraction, bulk Lorentz factor and jet power remain constant, the radius of the jet at which the synchrotron emission at frequency $\nu$ becomes optically thick is simply inversely proportional to the frequency $\nu$, $R\propto \nu^{-1}$.  

Let us now use this formula to estimate the radius of the transition region from the spectrum of J0531.  We shall use the physical parameters from Table \ref{tab1}, derived from the ballistic particle dominated jet fit to J0531 shown in Figures \ref{fig:1}b and \ref{fig:2}.   The observed frequency at which the synchrotron break occurs is approximately $3\times 10^{-5}\m{eV}=7.25\m{Ghz}$ and this corresponds to a plasma rest frame frequency $\nu_{\m{obs}}\approx 2\gamma_{\m{bulk}}\nu/(1+z)$.  In this case we find $R\approx 7.7\m{pc}$.  In our model using a scaled M87 jet geometry the radius of the transition region is $2000R_{s}$ at a distance $10^{5}R_{s}$ from the central black hole.  This places the transition region at a distance 380pc from the black hole corresponding to a black hole mass of $\approx 3.8\times 10^{10}M_{\odot}$.  This estimate roughly agrees with the inferred black hole mass of our fit, $2.5\times 10^{10}M_{\odot}$, and reinforces the conclusion that in the case of powerful, Compton-dominant blazars the majority of the synchrotron emission comes from relatively large distances from the central black hole.

\subsection{Frequency core-shift} 

We can use the general result in Equation \ref{knu2} to compare to observations of the frequency dependent core-shift in AGN jets.  Previous calculations of the core-shift have assumed a constant bulk Lorentz factor (\cite{1981ApJ...243..700K} and \cite{1998A&A...330...79L}).  Observations seem to show that generally $x_{\m{core}}\propto \nu_{\m{obs}}^{-k}$, with a mean value $k \approx1$ and a tail towards lower values of $k$ (\cite{2011A&A...532A..38S}).  We shall use our formulae to calculate the expected frequency core-shift in the case of the three conical jet models in Figure \ref{fig:1}a: the adiabatic equipartition jet, the ballistic equipartition jet and the ballistic particle dominated jet, and in the case of a parabolic accelerating jet.  Core-shift measurements generally probe frequencies from 1-40Ghz which correspond to large distances ($\approx 2000$pc for the fit to J0531) down to several tens of Swarzchild radii close to the base of the jet (43Ghz corresponds to $14-23R_{s}$ in M87 \cite{2011Natur.477..185H}).  With this large range of distances it is interesting to calculate the expected core-shift relation for both the outer conical and inner parabolic sections of the jet.  Assuming that the electron distribution remains fairly constant along the region of interest we use the formula.

\be
k_{\nu}d\m{x}\propto \frac{A_{\m{equi}}^{\frac{2+\alpha}{4}}}{(1+A_{\m{equi}})^{\frac{6+\alpha}{4}}}.\frac{W_{j}^{\frac{6+\alpha}{4}}d\m{x}}{R^{\frac{6+\alpha}{2}}\gamma_{\m{bulk}}\nu_{\m{obs}}^{\frac{4+\alpha}{2}}}. \label{kdx}
\ee

Let us assume that in the lab frame $R\propto x^{n}$ and the bulk Lorentz factor $\gamma_{\m{bulk}}\propto x^{m}$.  In this case the rest frame distance $d\m{x}$ over which the jet radius changes by a fixed fraction is proportional to $d\m{x}\propto x/\gamma_{\m{bulk}} \propto R^{1/n}/\gamma_{\m{bulk}}$.  To be compatible with the core-shift measurements ($x\propto \nu_{\m{obs}}^{-1}$), in general, requires that the equipartition fraction of the jet also depends on distance $A_{\m{equi}}\propto x^{w}$.  Let us also allow the remaining jet power in a section to change along the jet, $W_{j}\propto x^{u}$, due to adiabatic losses for example.  We can use Equation \ref{kdx} to calculate the expected core-shift properties of a blazar in the simplifying cases of $A_{\m{equi}} \gg1$, constant $A_{\m{equi}}$ and $A_{\m{equi}} \ll1$.

\be
k=\frac{2(4+\alpha)}{(6+\alpha)(2n-u)+4(w+2m-1)}, \qquad A_{\m{equi}}\gg1,
\ee

\be
k=\frac{2(4+\alpha)}{(6+\alpha)(2n-u)+4(2m-1)}, \qquad A_{\m{equi}}=\m{constant},
\ee

\be
k=\frac{2(4+\alpha)}{(6+\alpha)(2n-u)-(2+\alpha)w+8m-4}, \qquad A_{\m{equi}}\ll1.
\ee

To calculate the core-shift relation in the case of an accelerating parabolic jet ($n=0.5$, $m=0.5$) we assume that the jet is magnetically dominated close to the base of the jet so $A_{\m{equi}}\gg 1$, most of the jet power is not radiated away until larger distances, $u=0$, and $\alpha \approx 2$.  In order for the parabolic jet to produce the core-shift relation $k=1$ (compatible with the innermost part of the jet in M87 \cite{2011Natur.477..185H}), we require that $w=1$ which means the equipartition fraction increases along the parabolic section $A_{\m{equi}} \propto x$.  The jet plasma actually becomes more magnetically dominated further along the parabolic section and so contains a smaller proportion of its energy in non-thermal electrons.  Simulations find that jets start with a parabolic shape $R\propto x^{1/2}$ and an accelerating bulk Lorentz factor $\gamma_{\m{bulk}} \propto x^{1/2}$ (for example \cite{2006MNRAS.368.1561M}), so this result is interesting as this is the first time that the frequency core-shift has been calculated for a parabolic accelerating jet.  In our formula we have assumed that the jet is oriented close to our line of sight so the Doppler factor, $\delta_{\m{Doppler}} \propto \gamma_{\m{bulk}}$, however, this is only valid for blazars and needs to be modified for jets which are oriented at larger angles to our line of sight, such as M87. 

Let us now consider a conical ballistic jet ($n=1$, $m=0$) which starts particle dominated and becomes more particle dominated ($w=-1$) further along the jet.  Again we assume that the majority of the jet power is not radiated away until further along the jet $u=0$.  In this case we assume that $A_{\m{equi}}\ll1$. So in the case of the ballistic particle dominated model with $\alpha \approx2$, we calculate a core-shift relation $x\propto \nu_{\m{obs}}^{-0.75}$.  In the study by \cite{2011A&A...532A..38S} they found that the mean core-shift relation was $k=1$, however, this distribution is relatively wide with a median value of $k=1.2$ and a range $0.67\geq k\geq 2$ (with one outlying point at $k=0.29$), so it does not exclude this model.  Let us now consider the conical ballistic jet in equipartition ($n=1$, $w=0$) which suffers no electron energy losses ($u=0$) and has a constant bulk Lorentz factor ($m=0$).  The early calculation by \cite{1981ApJ...243..700K} found a relation $x\propto \nu^{-1}$ independent of $\alpha$ for this type of jet model, which is consistent with observations.  From Equation \ref{Rcon} we see that using our assumptions and $R\propto x$ we recover the well known $x\propto \nu_{\m{obs}}^{-1}$ relation.  

Finally, we consider the conical adiabatic equipartition jet with constant bulk Lorentz factor ($n=1$, $m=0$, $w=0$).  In this case the jet power decreases due to adiabatic losses on the electrons (since the jet is constantly in equipartition) so approximately $W_{j}\propto R^{-2/3}$ and $u=-2/3$.  For the case $\alpha \approx 2$ we find $k=0.69$, which is just within the range of values of $k$ found by \cite{2011A&A...532A..38S}.  The model which best fits the core-shift measurements is the simple ballistic conical jet in equipartition.  However, this model does not fit the radio slopes of these Compton-dominant blazars well, as shown in Figure \ref{fig:1}a.  Both the adiabatic equipartition jet and ballistic particle dominated jets fit the radio slope well but differ from the mean observed core-shift relation $x\propto \nu_{\m{obs}}^{-1}$.  The core-shift relations we have calculated for these models are both lower than the mean $k=1$ value with both models just within the observed distribution of $k$ from a sample of 20 objects from \cite{2011A&A...532A..38S}.  These results show that the frequency core-shift measurement is useful to distinguish between different jet models and geometries. 

\section{Fitting the multiwavelength spectra}

In the previous two subsections we have shown that the low-frequency observations indicate that the jet first comes into equipartition at large distances along the jet outside of the BLR and dusty torus.  The radio observations indicate that if the conical section of the jet is ballistic then it becomes particle dominated further along the jet and if the jet is adiabatic then the jet remains close to equipartition.  We shall now attempt to fit the six multiwavelength spectra using models in which the jet transitions from parabolic to conical at $10^{5}R_{s}$ with either a ballistic conical section which starts in equipartition at the transition region and becomes particle dominated further along the jet or an adiabatic conical jet in equipartition.

\subsection{The ballistic model}

\begin{figure*}
	\centering
		\subfloat[J0349]{ \includegraphics[width=8 cm, clip=true, trim=1cm 1cm 0cm 1cm]{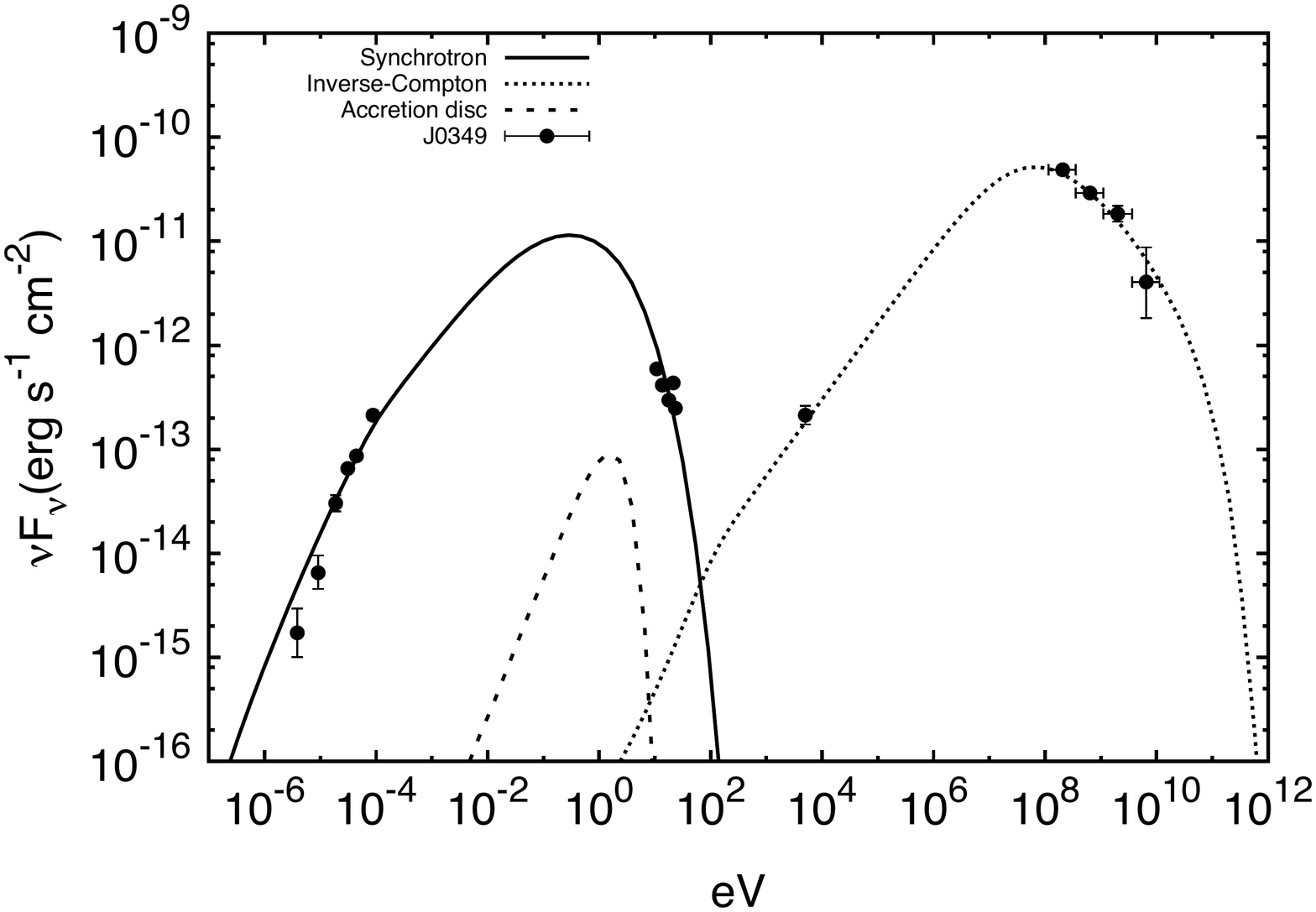} }
		\qquad
		\subfloat[J0457]{ \includegraphics[width=8cm, clip=true, trim=1cm 1cm 0cm 1cm]{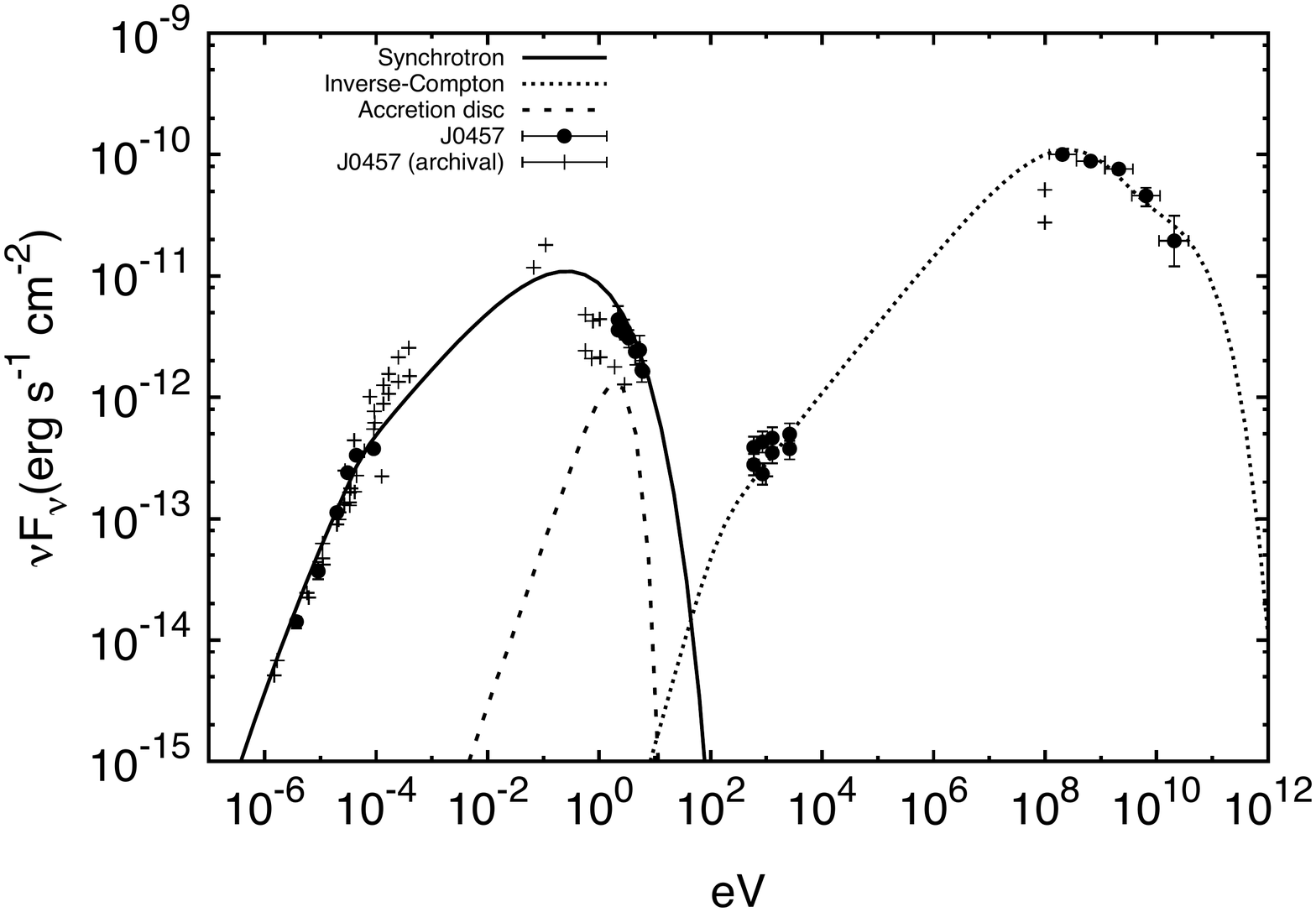} }
		\\
		\subfloat[J0531]{ \includegraphics[width=8cm, clip=true, trim=1cm 1cm 0cm 1cm]{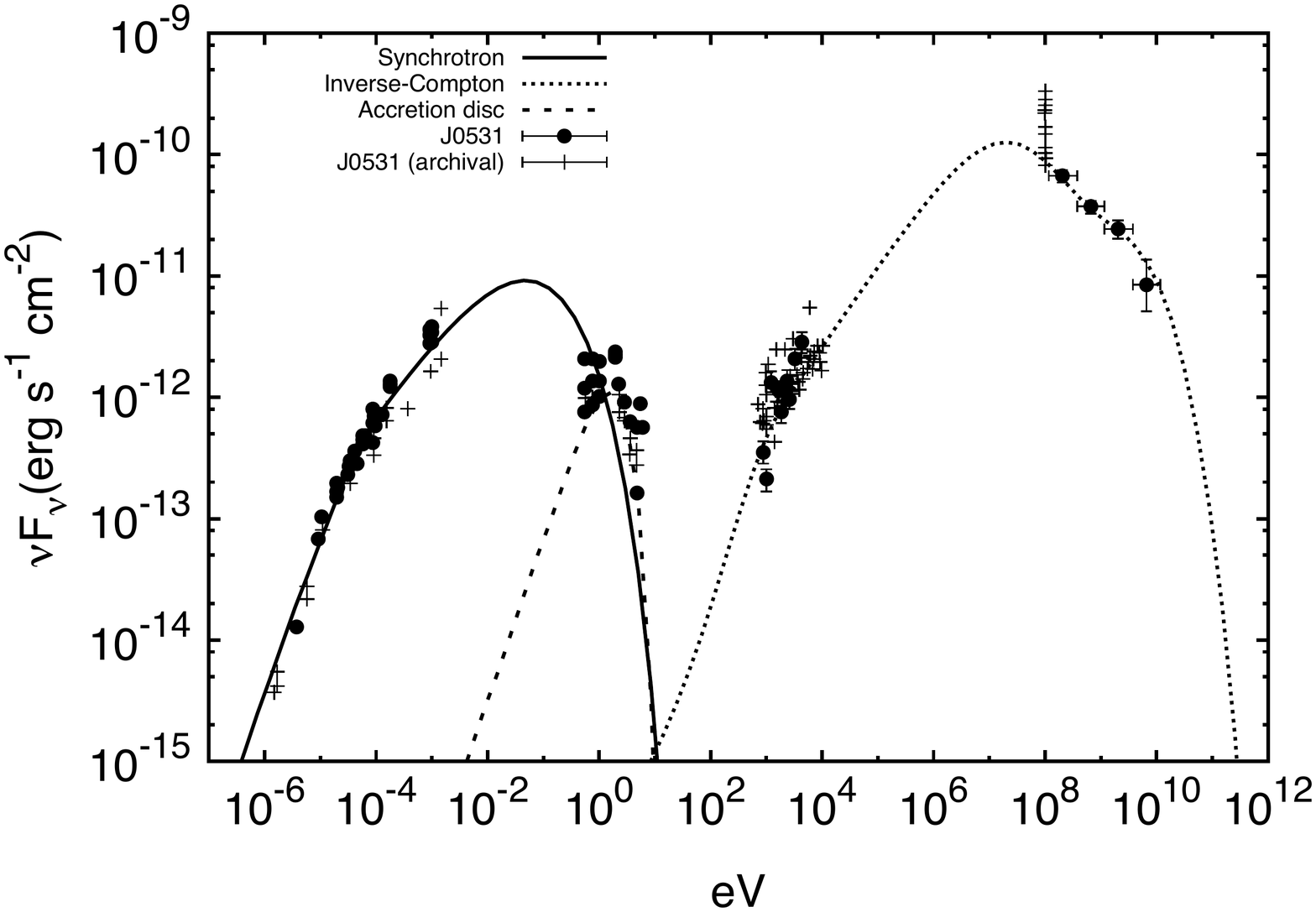} }
		\qquad
		\subfloat[J0730]{ \includegraphics[width=8 cm, clip=true, trim=1cm 1cm 0cm 1cm]{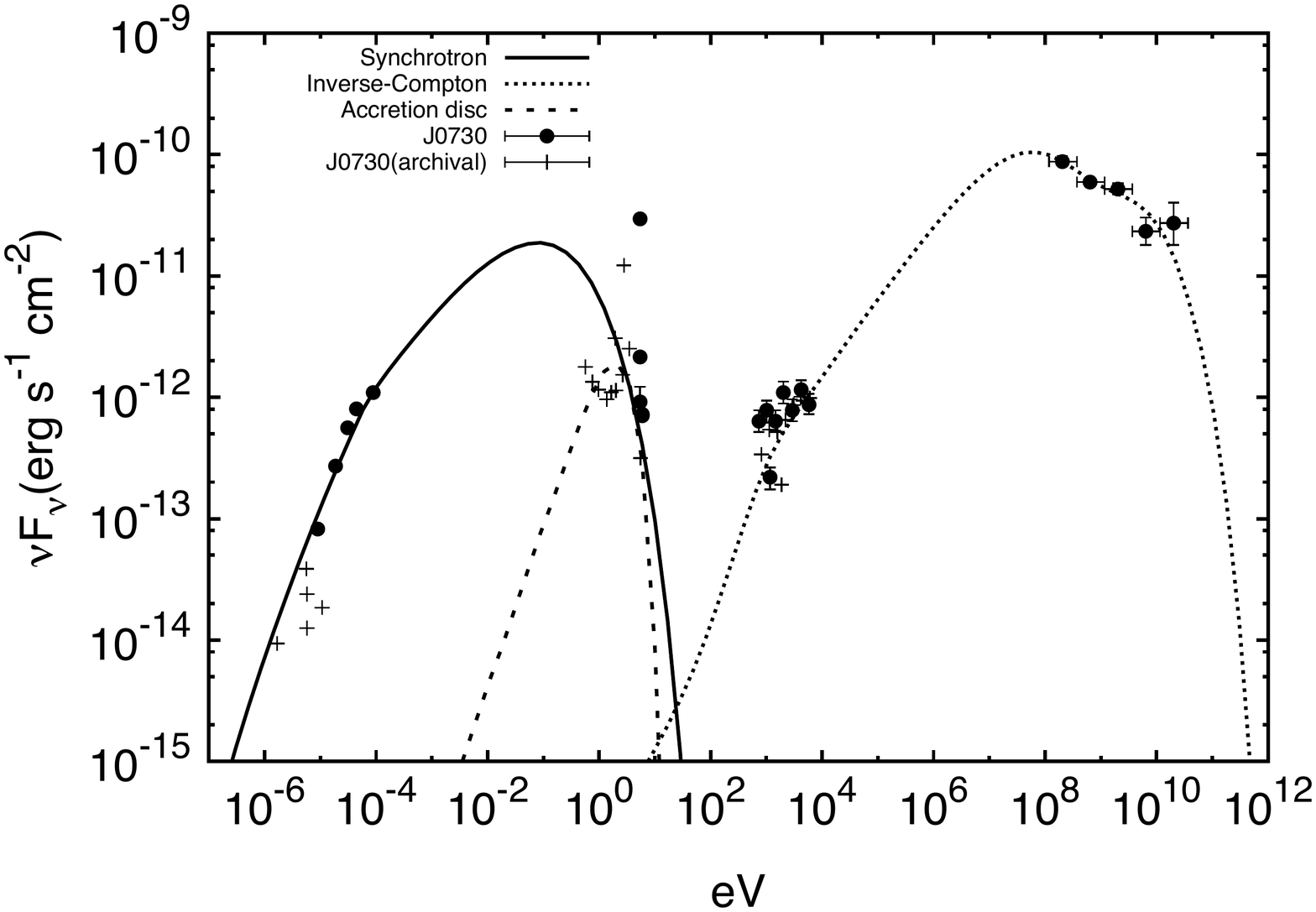} } 
		\\
		\subfloat[J1504]{ \includegraphics[width=8 cm, clip=true, trim=1cm 1cm 0cm 1cm]{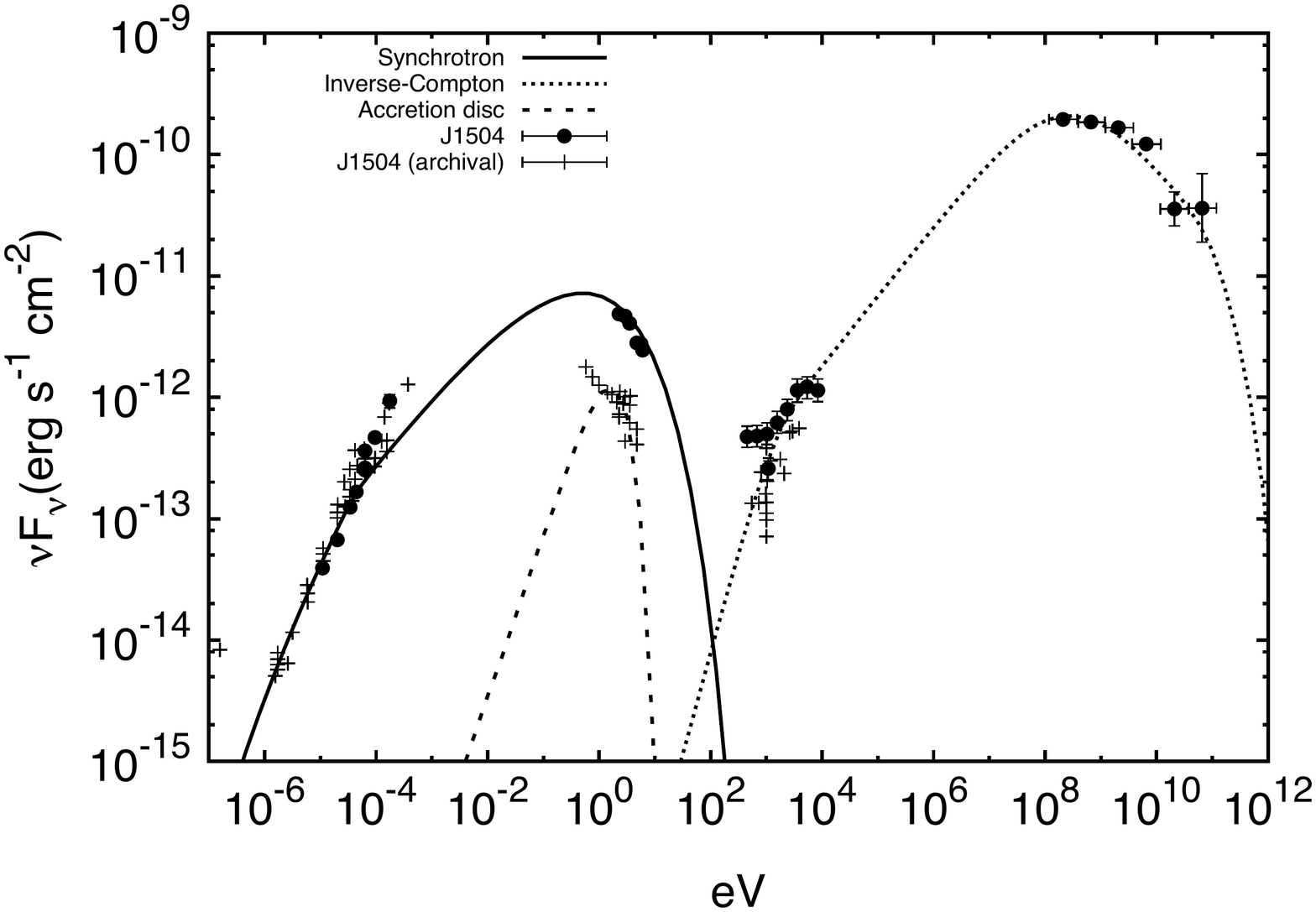} }
		\qquad
		\subfloat[J1522]{ \includegraphics[width=8 cm, clip=true, trim=1cm 1cm 0cm 1cm]{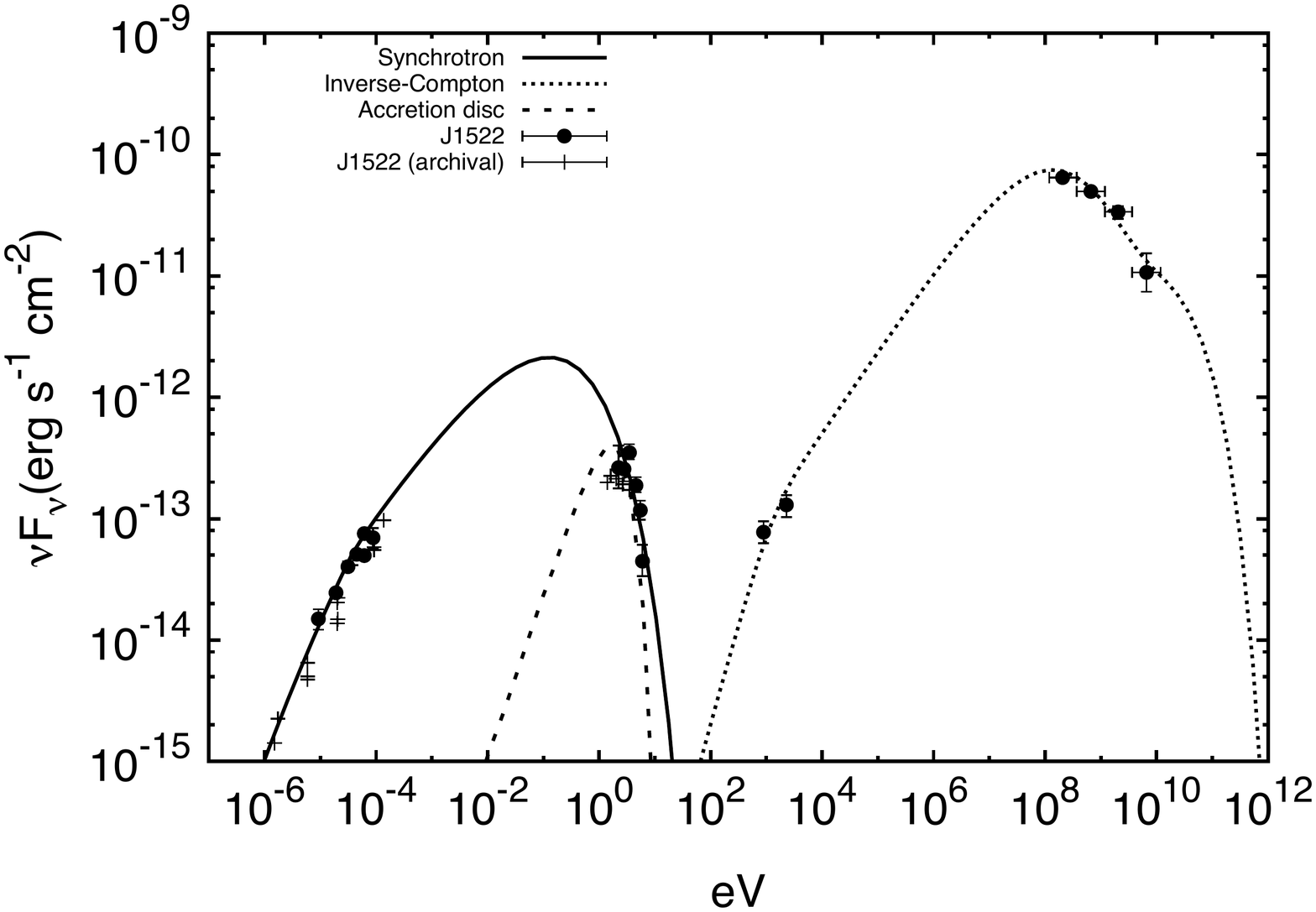} }
			
	\caption{This figure shows the results of fitting the ballistic jet model to the SEDs of J0349, J0457, J0531, J0730, J1504 and J1522. The model fits the observations very well for all six blazars across all wavelengths.  We find that the inverse-Compton emission of all the blazars is well described using CMB seed photons with a small contribution to the highest energy emission by scattering of NLR photons. }
	\label{fig:2}
\end{figure*}

\begin{figure*}
	\centering
		 \subfloat[J1504 ballistic]{ \includegraphics[width=8 cm, clip=true, trim=1cm 1cm 0cm 1cm]{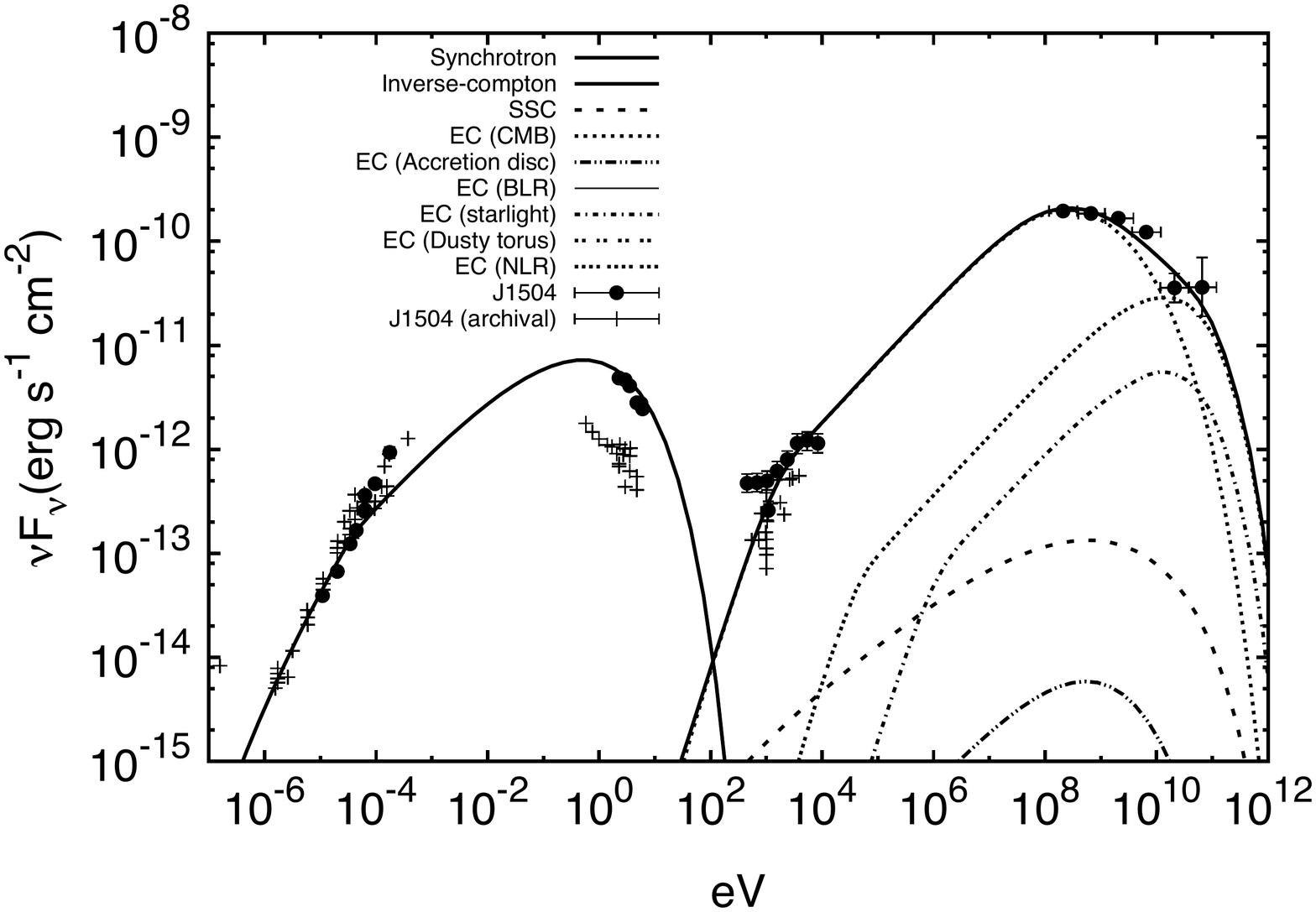} }
		\qquad
		\subfloat[J1504 adiabatic]{ \includegraphics[width=8cm, clip=true, trim=1cm 1cm 0cm 1cm]{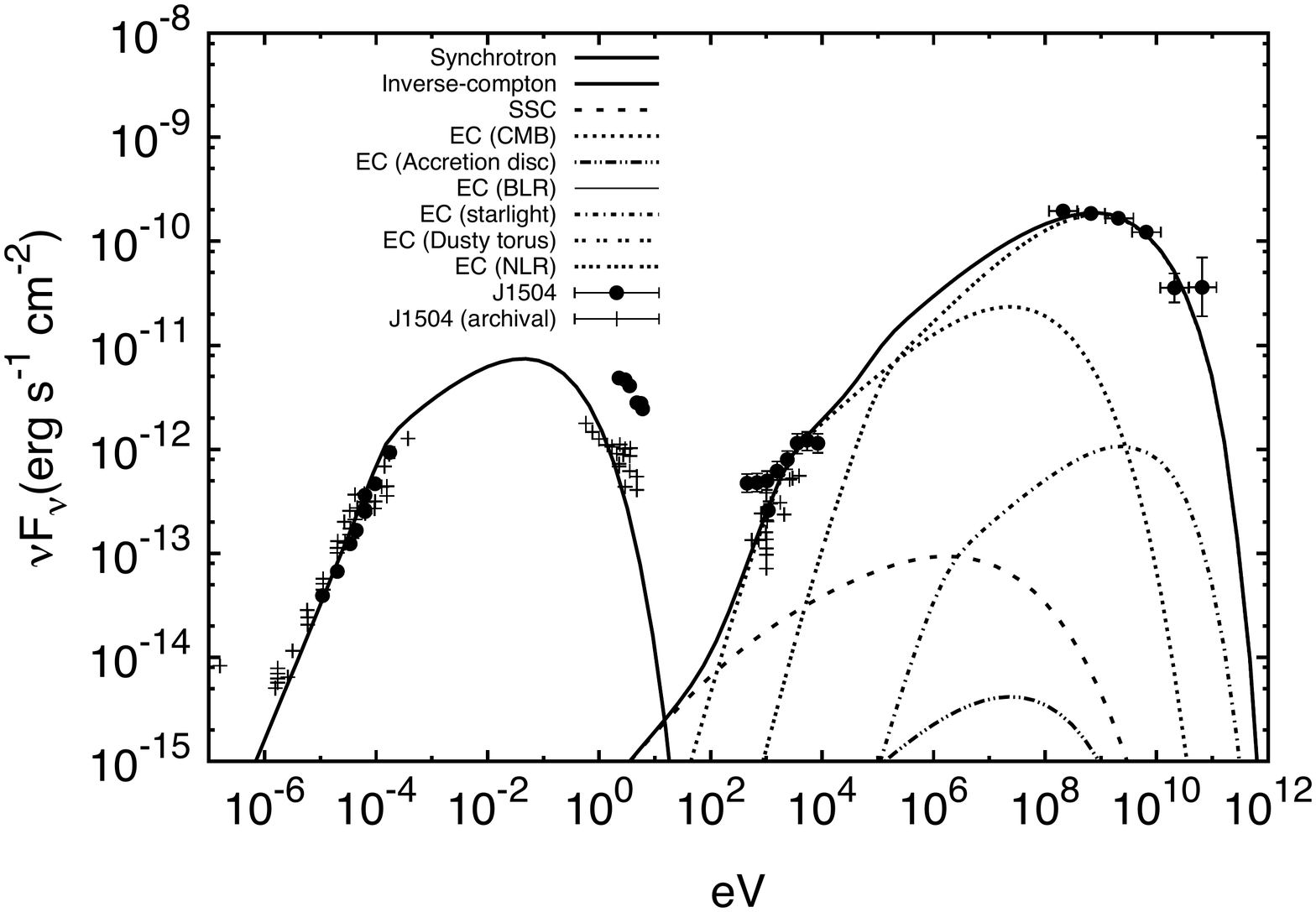} }
			
	\caption{This figure shows the different components of the inverse-Compton emission for the fit to J1504 for the two models.  Figure a shows that the inverse-Compton emission in the ballistic model is primarily due to scattering of CMB seed photons with a small contribution from NLR photons required to fit to the highest energy data point.  Figure b shows that in the adiabatic model the inverse-Compton emission is primarily due to scattering NLR photons with a contribution from scattering CMB photons at x-ray energies.  For these Compton-dominant blazars we find that the SSC emission is sub-dominant due to the comparitively low magnetic field strength at the transition region of the jet.}
	\label{fig:3}
\end{figure*}

\begin{table*}
\centering
\begin{tabular}{| c | c | c | c | c | c | c | c |}
\hline
Parameter & J0349.8 $-2102$ & J0457.1$-2325$ & J0531.0$+1331$ & J0730.4$-1142$ & J1504.4$+1030$ & J1522.2$+3143$ \\ \hline 
$W_{j}$ & $5.0 \times 10^{39}\rm{W}$ & $1.7 \times 10^{39}\rm{W}$ & $5.0 \times 10^{39}\rm{W}$ & $4.5 \times 10^{39}\rm{W}$ & $5.5 \times 10^{39}\rm{W}$ & $1.5 \times 10^{39}\rm{W}$  \\ \hline
L & $5 \times 10^{20}\rm{m}$ & $1 \times 10^{21}\rm{m}$ & $2 \times 10^{21}\rm{m}$ & $1 \times 10^{21}\rm{m}$ & $1 \times 10^{21}\rm{m}$ & $1 \times 10^{21}\rm{m}$ \\ \hline
$E_{\m{min}}$ & 5.11 \rm{MeV} & 5.11 \rm{MeV} & 15.3 \rm{MeV} & 15.3 \rm{MeV} & 15.3 \rm{MeV} & 15.3 \rm{MeV} \\ \hline
$E_{\m{max}}$ & 8.10 \rm{GeV} & 8.48 \rm{GeV} & 4.06 \rm{GeV} & 4.55 \rm{GeV} & 17.3\rm{GeV} & 5.48 \rm{GeV}  \\ \hline
$\alpha$ & 1.5  & 1.85  & 1.65 & 1.7 & 1.85 & 1.65  \\ \hline
$\theta'_{\m{opening}}$ & $3^{o}$ & $3^{o}$ & $3^{o}$ & $3^{o}$ & $3^{o}$ & $3^{o}$ \\ \hline
$\theta_{\m{observe}}$ & $1.0^{o}$ & $1.0^{o}$ & $1.0^{o}$ & $1.0^{o}$ & $1.0^{o}$ & $1.0^{o}$ \\ \hline
$\gamma_{\m{0}}$ & 4 & 4 & 4 & 4 & 4 & 4 \\ \hline
$\gamma_{\m{max}}$ & 30 & 35 & 25 & 28 & 45 & 40 \\ \hline
$\gamma_{\m{min}}$ & 15 & 20 & 15 & 15 & 20 & 20 \\ \hline
$M$ & $8.0 \times 10^{9}M_{\odot}$ & $1.4 \times 10^{10}M_{\odot}$ & $2.5 \times 10^{10}M_{\odot}$ & $2.0 \times 10^{10}M_{\odot}$ & $2.5 \times 10^{10}M_{\odot}$ & $1.3 \times 10^{10}M_{\odot}$ \\ \hline
$L_{\m{acc}}$ & $8.3 \times 10^{38}\rm{W}$ & $4.2 \times 10^{38}\rm{W}$ & $3.3 \times 10^{39}\rm{W}$ & $2.5 \times 10^{39}\rm{W}$ & $2.5 \times 10^{39}\rm{W}$ & $4.2 \times 10^{38}\rm{W}$ \\ \hline
$x_{\m{outer}}$ & $2\rm{kpc}$ & $2\rm{kpc}$ & $3\rm{kpc}$ & $1 \rm{kpc}$ & $1 \rm{kpc}$ & $1\rm{kpc}$ \\ \hline
$B_{T}$ & $2.65\times10^{-6}\rm{T}$ & $7.57\times10^{-7}\rm{T}$ & $1.02\times10^{-6}\rm{T}$ & $1.08\times10^{-6}\rm{T}$ & $5.93\times10^{-7}\rm{T}$ & $6.70\times10^{-7}\rm{T}$ \\ \hline
\end{tabular}
\caption{This table shows the values of the physical parameters used in the ballistic model fits shown in Figure $\ref{fig:2}$. We also include the value of the magnetic field strength in the plasma rest frame at the transition region for the fits ($B_{T}$).  }
\label{tab1}
\end{table*}

Figure $\ref{fig:2}$ shows the results of fitting the ballistic model to the six Compton-dominant FSRQs.  The model starts in equipartition and becomes particle dominated at larger distances along the jet.  The equipartition fraction changes as $U_{B}/U_{e}=2x_{T}/(x_{T}+x)$ for $x_{T} \geq x \geq 100 x_{T}$ and $U_{B}/U_{e}=1/50$ for $x >100x_{T}$, where $x_{T}$ is the distance of the transition region along the jet in the lab frame.  The model fits very well to the observed spectra across all wavelengths.  We show the physical parameters of the fits to the spectra in Table $\ref{tab1}$.  We find that for all six blazars their inverse-Compton emission and Compton-dominance is well fitted by scattering of Doppler-boosted CMB photons.  The different components contributing to the inverse-Compton emission are shown in Figure \ref{fig:3}a.  This is a very interesting result since it is widely believed that Compton-dominance is due to inverse-Compton scattering of BLR scattered photons (\cite{2008MNRAS.391.1981M}).  This is the first time that a realistic extended jet model has been used to fit to the spectrum of a sample of blazars and these new results show that it is important to consider the jet as a whole, including radio observations, to determine the physics governing these objects.

We find consistently for all these blazars that fits to their spectra require high power, high bulk Lorentz factor jets observed close to the line of sight.  This is consistent with our expectations from the blazar sequence (in radio power) and from AGN unification since we expect Compton-dominant blazars to be high power, high bulk Lorentz factor FRII-type jets observed close to the line of sight.  We also find that for these objects the black hole mass (which completely determines the geometry of the parabolic region and the transition region) that is inferred from the fits to the spectra are all large $>10^{9}M_{\odot}$.  This is consistent with Compton-dominant blazars being large elliptical galaxies with powerful FRII type jets.   

Interestingly, the black hole mass inferred from the distance of the transition region agrees with that inferred independently from fitting the accretion disc spectrum to J0531 only if the transition region occurs at $10^{5}R_{s}$ as in the jet of M87.  This supports the idea that the transition region occurs at $10^{5}R_{s}$ in jets and scales linearly with black hole mass.  

\subsection{The adiabatic model}

\begin{figure*}
	\centering
		\subfloat[J0349]{ \includegraphics[width=8 cm, clip=true, trim=1cm 1cm 0cm 1cm]{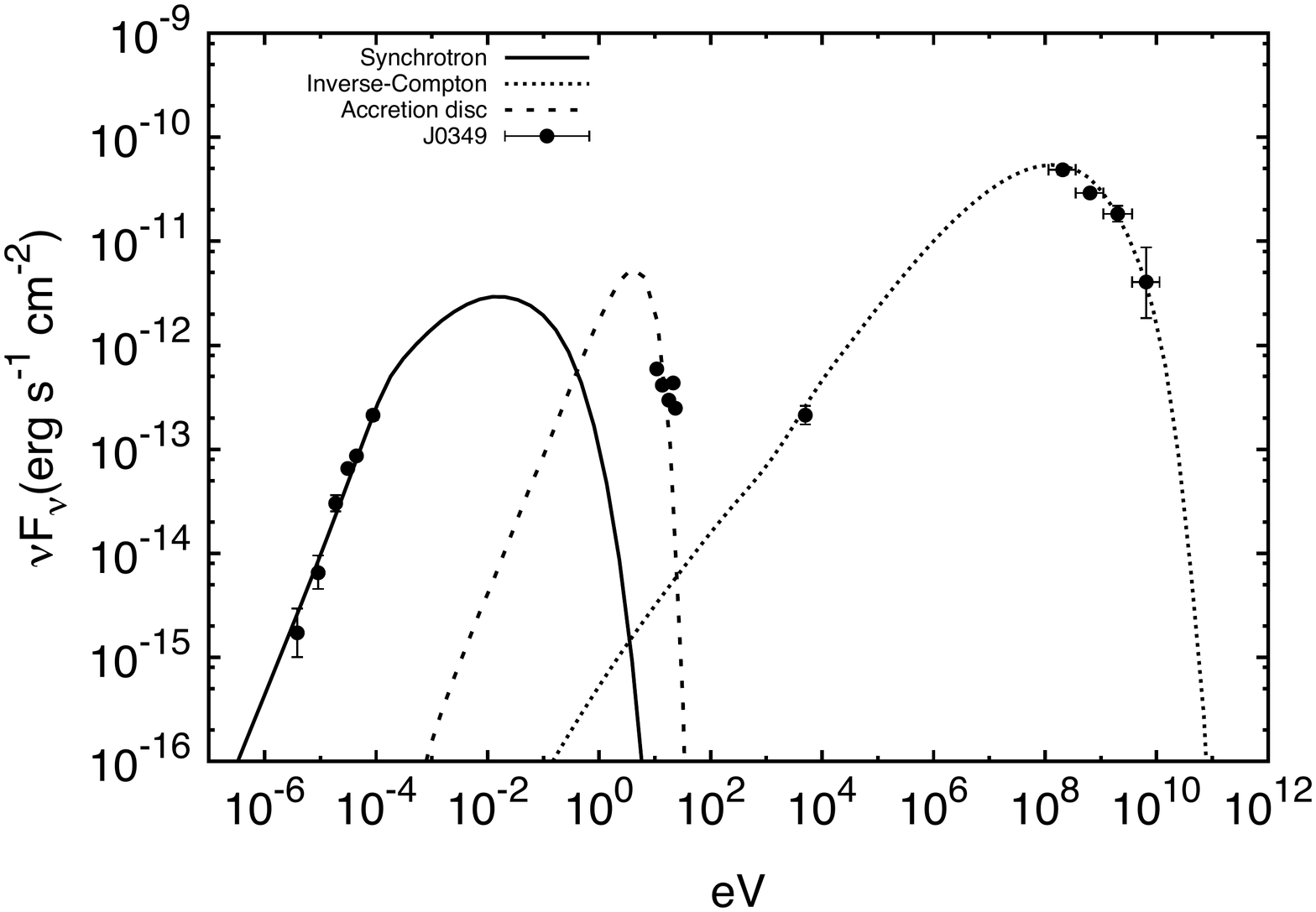} }
		\qquad
		\subfloat[J0457]{ \includegraphics[width=8cm, clip=true, trim=1cm 1cm 0cm 1cm]{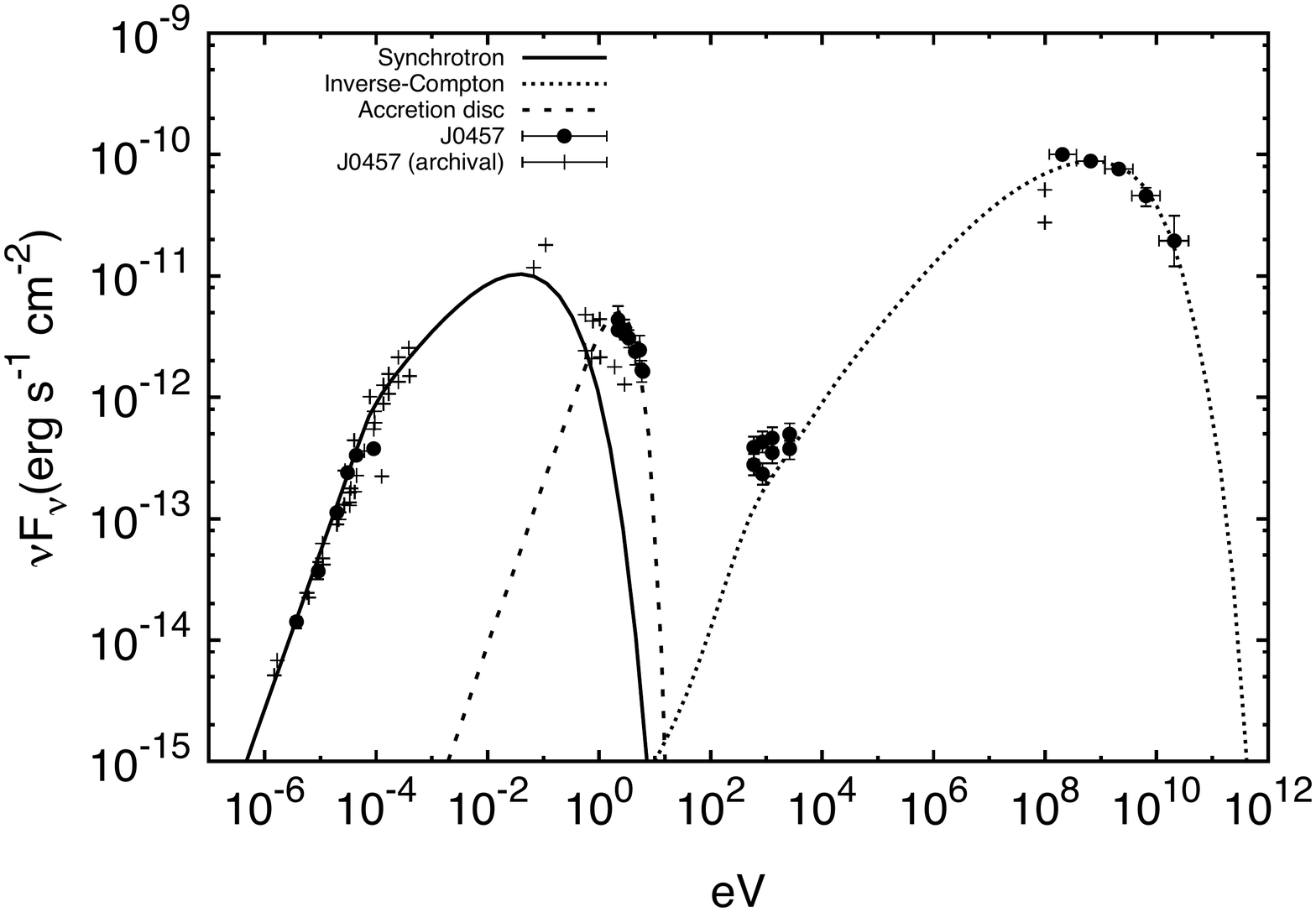} }
		\\
		\subfloat[J0531]{ \includegraphics[width=8cm, clip=true, trim=1cm 1cm 0cm 1cm]{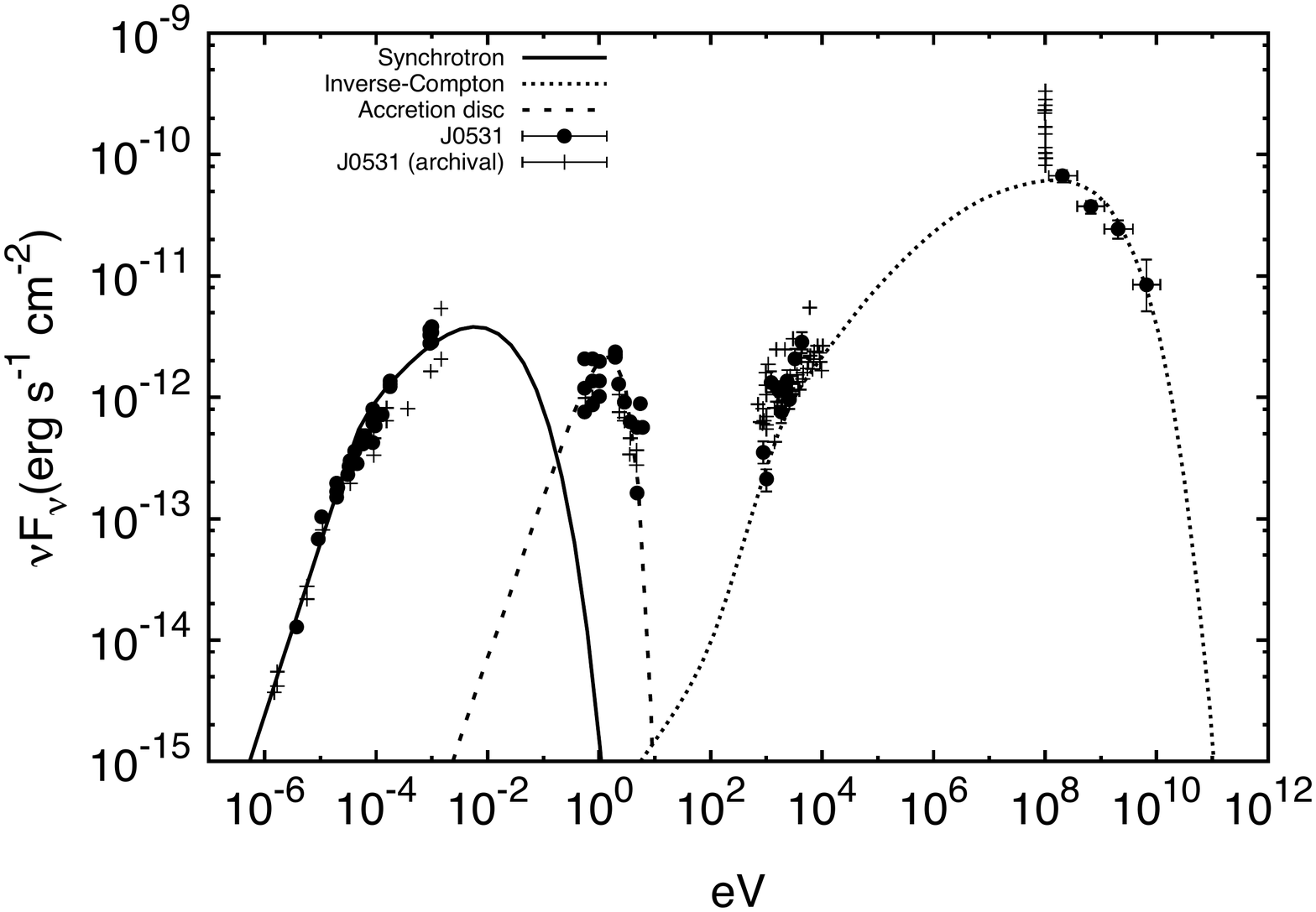} }
		\qquad
		\subfloat[J0730]{ \includegraphics[width=8 cm, clip=true, trim=1cm 1cm 0cm 1cm]{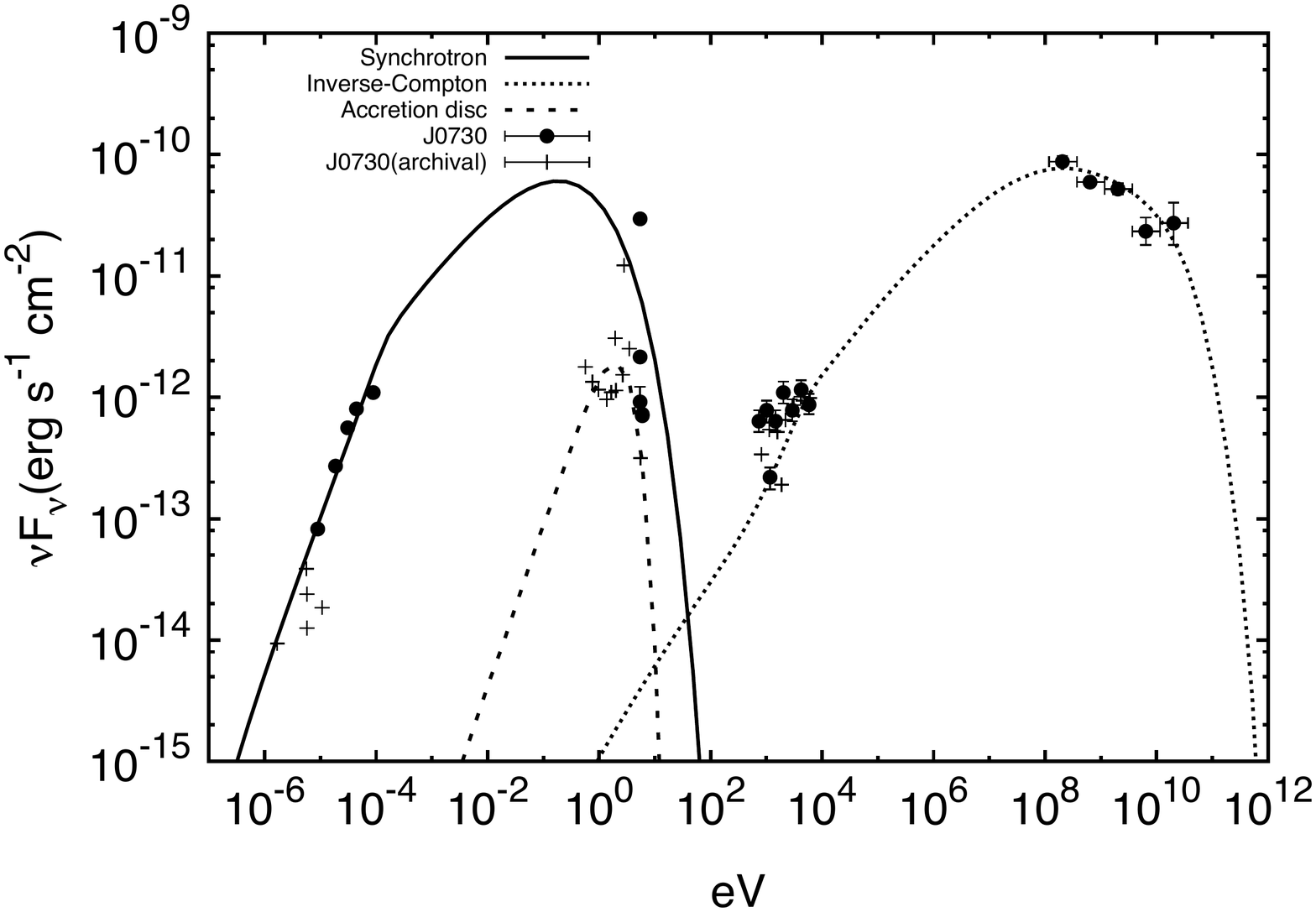} } 
		\\
		\subfloat[J1504]{ \includegraphics[width=8 cm, clip=true, trim=1cm 1cm 0cm 1cm]{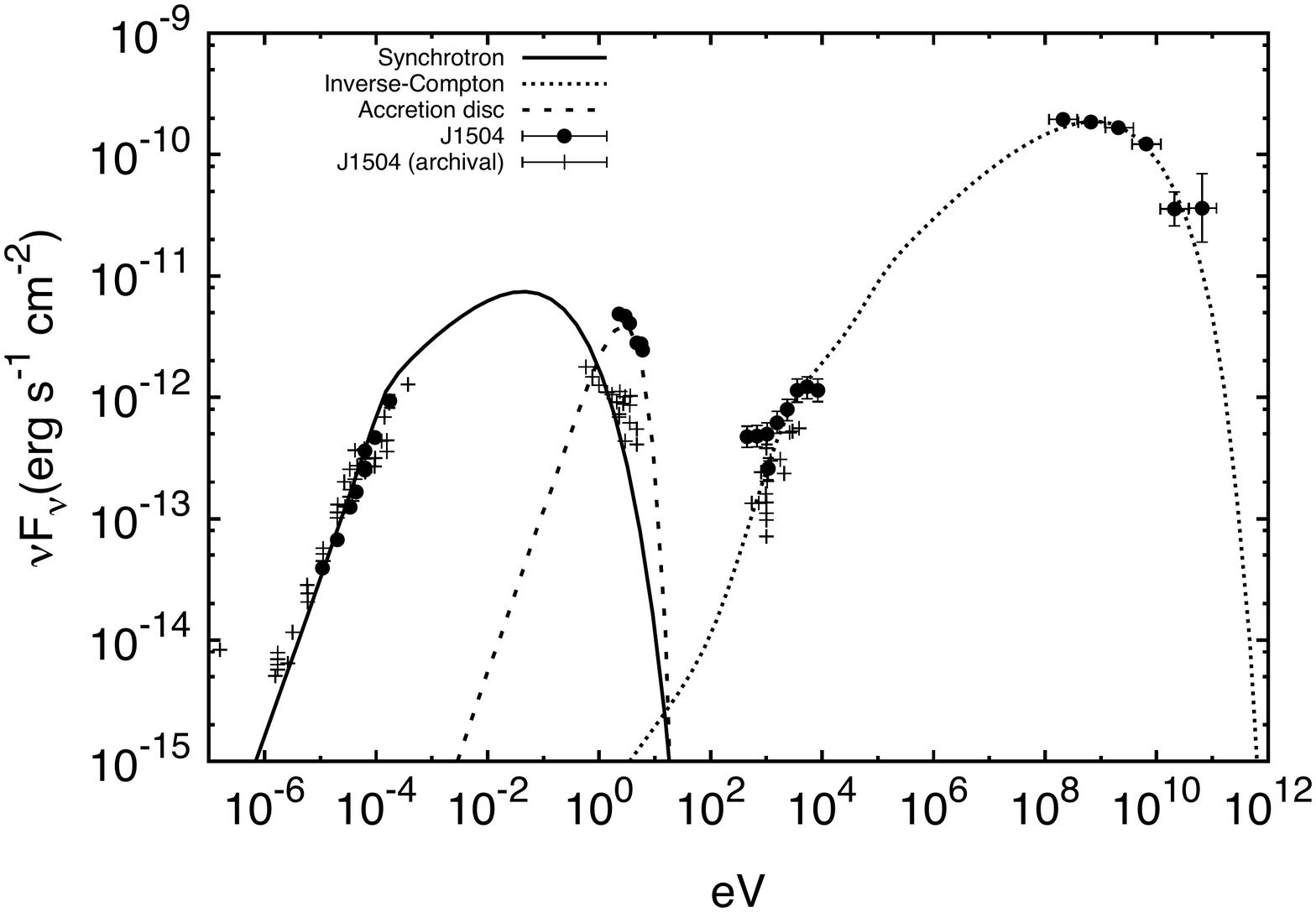} }
		\qquad
		\subfloat[J1522]{ \includegraphics[width=8 cm, clip=true, trim=1cm 1cm 0cm 1cm]{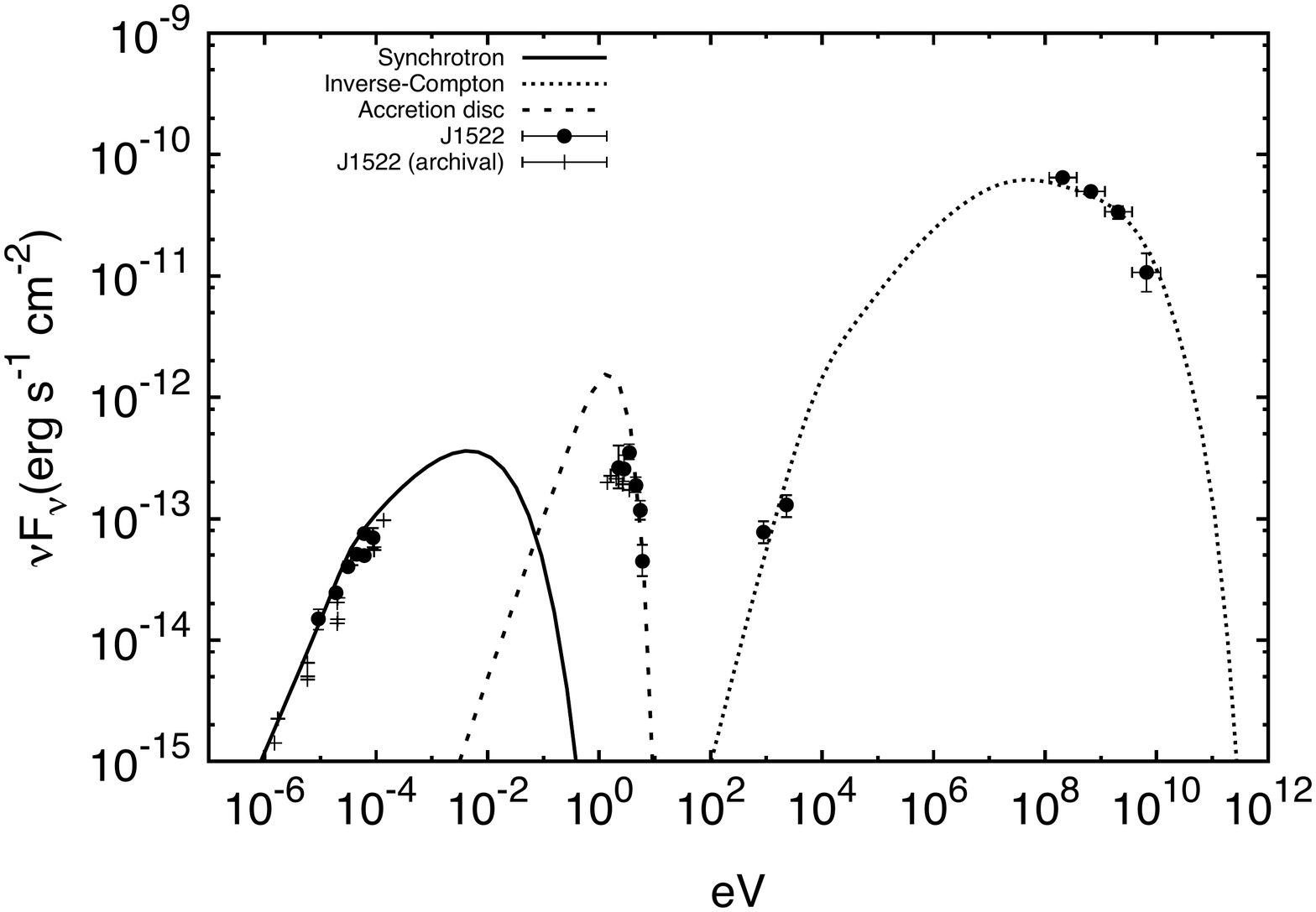} }
			
	\caption{This figure shows the results of fitting the adiabatic jet model to the SEDs of J0349, J0457, J0531, J0730, J1504 and J1522. The model fits the observations well across all wavelengths, although for some of the blazars the model requires very high bulk Lorentz factors ($\gamma_{\m{bulk}}=100$ for J1522).  We find that the inverse-Compton emission of all the blazars is well described using NLR seed photons with a contribution to the x-ray emission from scattering of CMB photons. }
	\label{fig:4}
\end{figure*}

Figure \ref{fig:4} shows the results of fitting the adiabatic jet model in equipartition to the observations of the six Compton-dominant blazars.  The model fits the data well across all wavelengths, however, to fit the Compton-dominant spectra requires very high bulk Lorentz factors as shown in Table \ref{tab2}.  The high bulk Lorentz factors are required in order that the rate of inverse-Compton emission from scattering CMB and NLR photons is not much smaller than the rate of adiabatic losses.  In the case of J1522 we find that the adiabatic equipartition model struggles to reproduce the Compton-dominance of the spectrum and requires an extreme value for the bulk Lorentz factor at the transition region $\gamma_{\m{bulk}}=100$, this is much higher than the expected value of the bulk Lorentz factor in blazars inferred from superluminal motion \cite{2001ApJS..134..181J}.  For this reason and the low value of $k$ calculated in Section 5.1 for the frequency core-shift we disfavour the adiabatic equipartition jet.  

We find that the fits all require high power, high bulk Lorentz factor jets observed close to the line of sight as we expect for Compton-dominant blazars and in agreement with the ballistic fits.  We find that all the fits have transition regions at large distances outside the BLR and dusty torus with large inferred black hole masses $>10^{9}M_{\odot}$.  The inverse-Compton emission is due primarily to the scattering of NLR and CMB photons at large distances along the jet.  We show the different components of the inverse-Compton emission for the fit to J1504 in Figure \ref{fig:3}b.  As for the ballistic model, the black hole mass inferred from the distance of the transition region agrees with that inferred independently from the accretion disc spectrum of J0531 if the transition region occurs at $10^{5}R_{s}$ as in the jet of M87.  

We have also considered an adiabatic model in which the jet is in equipartition after the transition region but with some electron acceleration occuring at smaller distances along the jet, in order to produce the gamma-rays and help reduce the bulk Lorentz factor.  In this case we find that accelerating electrons within the dusty torus tends to produce too much synchrotron emission due to the relatively high energy density in the magnetic field compared to the external photon field (both measured in the plasma rest frame) resulting in the overproduction of synchrotron emission.  Accelerating electrons in the BLR also suffers from a similar problem unless the location of the acceleration is fine-tuned to be close to the outer radius of the BLR.  This is because  the energy density in the BLR is relatively constant with distance $x$, whilst $U_{B} \propto x^{-2}$. So the Compton-dominance, which is approximately proportional to $U_{\gamma}/U_{B}$, increases towards the outer radius of the BLR.  In an accelerating jet model the plasma has a lower bulk Lorentz factor at small distances within parabolic region and BLR, so the emission is not so strongly Doppler-boosted as that from further along the jet close to or beyond the transition region.  We do not find that accelerating electrons within the BLR in the quiescent state is able to reproduce the observed data well and requires fine-tuning of the acceleration region within the BLR.  For these reasons we do not favour this scenario in order to explain the quiescent spectra. 

\begin{table*}
\centering
\begin{tabular}{| c | c | c | c | c | c | c | c |}
\hline
Parameter & J0349.8 $-2102$ & J0457.1$-2325$ & J0531.0$+1331$ & J0730.4$-1142$ & J1504.4$+1030$ & J1522.2$+3143$ \\ \hline 
$W_{j}$ & $8.0 \times 10^{39}\rm{W}$ & $8.0 \times 10^{39}\rm{W}$ & $1.4 \times 10^{40}\rm{W}$ & $3.0 \times 10^{40}\rm{W}$ & $2.0 \times 10^{40}\rm{W}$ & $8.0 \times 10^{39}\rm{W}$  \\ \hline
L & $1 \times 10^{21}\rm{m}$ & $2 \times 10^{21}\rm{m}$ & $2 \times 10^{21}\rm{m}$ & $1 \times 10^{21}\rm{m}$ & $2 \times 10^{21}\rm{m}$ & $2 \times 10^{21}\rm{m}$ \\ \hline
$E_{\m{min}}$ & 5.11 \rm{MeV} & 5.11 \rm{MeV} & 10.2 \rm{MeV} & 15.3 \rm{MeV} & 10.2 \rm{MeV} & 15.3 \rm{MeV} \\ \hline
$E_{\m{max}}$ & 1.28 \rm{GeV} & 2.03 \rm{GeV} & 1.14 \rm{GeV} & 4.55 \rm{GeV} & 2.87 \rm{GeV} & 1.09 \rm{GeV}  \\ \hline
$\alpha$ & 1.3  & 1.6  & 1.55 & 1.7 & 1.9 & 1.5  \\ \hline
$\theta'_{\m{opening}}$ & $3^{o}$ & $3^{o}$ & $3^{o}$ & $3^{o}$ & $3^{o}$ & $3^{o}$ \\ \hline
$\theta_{\m{observe}}$ & $1.25^{o}$ & $1.0^{o}$ & $0.8^{o}$ & $1.0^{o}$ & $0.8^{o}$ & $0.6^{o}$ \\ \hline
$\gamma_{\m{0}}$ & 4 & 4 & 4 & 4 & 4 & 4 \\ \hline
$\gamma_{\m{max}}$ & 25 & 50 & 40 & 50 & 60 & 100 \\ \hline
$\gamma_{\m{min}}$ & 10 & 30 & 30 & 30 & 30 & 30 \\ \hline
$M$ & $8.0 \times 10^{9}M_{\odot}$ & $2.0 \times 10^{10}M_{\odot}$ & $4.2 \times 10^{10}M_{\odot}$ & $2.0 \times 10^{10}M_{\odot}$ & $1.5 \times 10^{10}M_{\odot}$ & $3.0 \times 10^{10}M_{\odot}$ \\ \hline
$L_{\m{acc}}$ & $5.0 \times 10^{40}\rm{W}$ & $1.7 \times 10^{39}\rm{W}$ & $6.8 \times 10^{39}\rm{W}$ & $2.5 \times 10^{39}\rm{W}$ & $8.3 \times 10^{39}\rm{W}$ & $1.7 \times 10^{39}\rm{W}$ \\ \hline
$x_{\m{outer}}$ & $1\rm{kpc}$ & $1\rm{kpc}$ & $1\rm{kpc}$ & $1 \rm{kpc}$ & $1 \rm{kpc}$ & $1\rm{kpc}$ \\ \hline
$B_{T}$ & $4.02\times10^{-6}\rm{T}$ & $8.05\times10^{-7}\rm{T}$ & $6.34\times10^{-7}\rm{T}$ & $1.56\times10^{-6}\rm{T}$ & $1.41\times10^{-6}\rm{T}$ & $2.68\times10^{-7}\rm{T}$ \\ \hline
\end{tabular}
\caption{This table shows the values of the physical parameters used in the adiabatic equipartition model fits shown in Figure $\ref{fig:3}$. We also include the value of the magnetic field strength in the plasma rest frame at the transition region for the fits ($B_{T}$).  }
\label{tab2}
\end{table*}

\subsection{Inverse-Compton scattering of external photons}

Our model uses a jet geometry based on radio observations of the jet in M87 scaled linearly with black hole mass.  For large black hole masses ($>10^{9}M_{\odot}$) the distance of the transition region of the jet, where the jet is expected to be close to equipartition, is $>10$\,pc.  This is consistent with our finding that BLR photons do not contribute significantly to the inverse-Compton emission of these Compton-dominant blazars since the BLR region is expected to be $<1$\,pc (\cite{2005ApJ...629...61K}).  This means that the synchrotron bright transition region of our jet is outside the BLR and so the BLR photons will illuminate the jet from behind and be Doppler-deboosted.  We expect the transition region of the jet to dominate the high energy emission because the plasma is in equipartition and has the highest bulk Lorentz factor at this point giving the largest Doppler-boosting of the emission.  From a reversal of the minimum energy argument (\cite{1959ApJ...129..849B}) we know that for a given total energy, a plasma emits the most synchrotron power when the plasma is close to equipartition, which is found to be near to the transition region in jet simulations where the jet transitions from parabolic to conical (\cite{2006MNRAS.368.1561M} and \cite{2009MNRAS.394.1182K}).

We show the different components of the inverse-Compton emission for the ballistic jet fit to blazar J1504 in Figure $\ref{fig:3}$.  The figure shows that the inverse-Compton emission is dominated by scattering of CMB photons with a small contribution from scattering NLR photons at high energies.  For a jet with a transition region outside the BLR and dusty torus we expect CMB and NLR photons to dominate the external photon field (see Figure 4 in Paper II).  At low redshifts we expect the energy density in starlight at the centre of an elliptical galaxy to dominate CMB photons (\cite{2011MNRAS.415..133H}).  However, these Compton-dominant objects are observed at relatively high redshifts so the CMB energy density is larger since $\rho_{\m{CMB}} \propto (1+z)^{4}$.  So we expect the CMB to be the dominant scattered photon field several kiloparsecs from the central black hole.  This is interesting since Compton-dominant blazars are observed predominantly at high redshift, which is exactly what we would expect if their inverse-Compton emission were due to scattering CMB photons.  Our model predicts that as we go to low redshifts where the energy density of the CMB decreases, high power, high bulk Lorentz factor blazars should become less Compton-dominant than an equivalent blazar at high redshift.  We wish to investigate whether this prediction for the Compton-dominance of low redshift blazars holds in future work. 

In order for an electron distribution to produce a Compton-dominant spectrum from SSC emission we require that the synchrotron energy density is so high that a \lq\lq{}Compton-catastrophe\rq\rq{} occurs.  This happens when electrons emit so much synchrotron radiation that the synchrotron radiation field is dense enough that scattering of photons is frequent and electrons lose more energy through inverse-Compton scattering than synchrotron emission.  In a jet with a fixed total energy and equipartition fraction we require a small jet radius at the transition region in order for the synchrotron radiation field to be dense enough for a \lq\lq{}Compton-catastrophe\rq\rq{} to occur. We find that in such small radius/large magnetic fields the synchrotron radiation emitted by the highest energy electrons is much higher than the observed energy of the synchrotron peak frequency and so we are unable to fit to the observed spectra using SSC emission.

\subsection{Acceleration of electrons along the jet}

We find that the electron distribution power law indices are lower than expected from parallel shock acceleration ($\alpha=2$) for these Compton-dominant blazars.  This suggests acceleration by oblique shocks (\cite{2011MNRAS.tmp.1506B} and \cite{2012ApJ...745...63S}), possibly a stationary recollimation shock at the transition region or acceleration by magnetic reconnection (\cite{2001ApJ...562L..63Z} and \cite{2007ApJ...670..702Z}).  We see clear evidence for this in the spectrum of the blazar J0531 where the synchrotron emission starts self-absorbed and close to flat and retains a steep gradient when the spectrum becomes optically thin (see Figure $\ref{fig:4}$b).  This optically thin spectrum requires a steep electron spectral index to fit to the observed emission.  In all these objects we also see that the inverse-Compton emission rises steeply from low energies to its peak.  This steep peak is also consistent with a low electron spectral index.  This is an interesting result since we found in Paper I that the spectrum of BL Lacertae was consistent with an electron spectral index $\alpha=2$, compatible with a parallel shock.  We wish to investigate whether other BL Lac type objects require spectral indices that are higher than those of Compton-dominant objects. This could imply a difference in the electron acceleration mechanism in different blazars possibly due to the different powers and bulk Lorentz factors of their jets.

\subsection{Compton drag}

In this investigation we have not explicitly calculated the effect of Compton drag on the jet plasma since we do not know the additional amount of deceleration caused by the entrainment of matter surrounding the jet. We have used the simple prescription for the deceleration of the jet set out in Equation 5 of Paper II and fitted the deceleration parameters for a spectrum. We find for both the ballistic and adiabatic fits the majority of the jet\rq{}s initial energy is not radiated away through anisotropic scattering of external photons and so the effect of Compton drag on the jet plasma is comparable to or smaller than half of the total amount of deceleration in the fits ($\Delta \gamma_{\m{bulk}} < 0.2\gamma_{\m{max}}$). In the case of BL Lacs whose emission is dominated by synchrotron and SSC emission we do not expect Compton drag to be a significant contribution to the jet deceleration. This is because synchrotron and SSC emission from an isotropic electron distribution in a small scale tangled field are emitted isotropically in the plasma rest frame and so do not result in deceleration of the jet. This means that we are justified in not explicitly calculating the deceleration due to Compton drag in these Compton-dominant FSRQs or BL Lacs since the contribution of Compton drag to the total deceleration of the jet is comparable to or smaller than that of entrainment.

\subsection{The blazar sequence}

We find that the slowly decelerating conical region, which is responsible for the radio emission, produces a nearly flat radio spectrum which is consistent with observations.  The frequency at which the observed synchrotron spectrum transitions from optically thick to optically thin emission is governed chiefly by the black hole mass which sets the radius of the bright transition region (the base of the conical region) and largely determines the magnetic field strength at this point.  The transition from optically thick to optically thin synchrotron emission at these comparatively low energies (as shown in the spectra of J0457 and J0531) implies large black hole masses for these blazars.  This is because a transition to optically thin synchrotron emission at low frequencies implies a low magnetic field strength at the base of the conical region and therefore a large radius as shown in Equation \ref{Rcon}. 

The base of the conical region in our model is in equipartition and so is responsible for producing the highest energy optically thin synchrotron emission.  The peak frequency of the synchrotron emission is then governed by the Doppler factor, the magnetic field strength at the base of the conical region and the maximum electron energy.  We find that the low synchrotron peak frequency of these Compton-dominant blazars is due to their large black hole masses which result in the transition region occuring at large distances from the black hole and with large radii.  These large radii result in relatively low magnetic field strengths at the bright transition region which produce the low peak frequency synchrotron emission.  We therefore expect that low powered BL Lac objects with high synchrotron peak frequencies should have lower black hole masses than low synchrotron peak frequency objects such as Compton-dominant objects.  

This can be illustrated through the following simple calculation.  We assume that the base of the conical region at a distance $10^{5}r_{s}$ from the central black hole with a radius $2000 r_{s}$ is in equipartition and has the maximum bulk Lorentz factor $\gamma_{\m{max}}$, so we expect it to dominate the optically thin synchrotron emission.  For an electron distribution with a maximum Lorentz factor $\gamma_{e}$ and an initial kinetic Luminosity $W_{j}$ using the formula for the synchrotron critical frequency (Equations \ref{knu} and \ref{jnu}) and Equation $\ref{BT}$ we find

\be
U_{B}=\frac{B_{T}^{2}\pi R_{T}^{2}}{2\mu_{0}} \times \frac{4\gamma_{max}^{2}}{3}\approx \frac{W_{j}}{2 c}, \qquad R_{T}=2000 \frac{GM}{c^{2}},
\label{BT}
\ee

\be
\nu_{\m{peak}}=\frac{3\delta_{\m{Dopp}}eE_{\m{max}}^{2}}{4\pi(1+z) m_{e}^{3}c^{4}}. \sqrt{\dfrac{3W_{j}\mu_{0}}{4\gamma_{\m{max}}^{2}\pi c (2000GM/c^{2})^{2}}},
\ee
\be
\nu_{\m{peak}} \propto W_{j}^{1/2}/M_{BH},
\ee

where a subscript T denotes the value of a quantity at the transition region, $M_{BH}$ is the central black hole mass, $\beta$ is the velocity of the jet divided by c, $\theta_{\m{obs}}$ is the observation angle to the jet axis, $z$ is the redshift and $\delta_{\m{Dopp}}$ is the doppler factor for the jet.  We have assumed that half the initial jet power is contained in magnetic field at the transition region where the plasma is in equipartition.  We have also assumed the plasma is described by a relativistic perfect fluid in its rest frame.  From this we see that for objects which differ only by black hole mass $\nu_{\m{peak}}\propto 1/M_{BH}$.  If we further assume that in general the jet power is a fixed fraction of the Eddington luminosity then $W_{j}\propto M_{BH}$ and so $\nu_{\m{peak}}\propto M_{BH}^{-1/2}$. If blazar jets are reasonably well described by a scaled M87 geometry this could even be used as a rough estimate of the black hole mass of a blazar based on its synchrotron peak frequency. 

It would be an exciting result if the black hole mass of a blazar largely determines the peak frequency of its synchrotron emission. It is perhaps not surprising if in general FSRQs have higher black hole masses than BL Lacs.  If jet power is fundamentally related to accretion, as currently believed, then we would expect powerful FSRQs to be accreting at a higher rate, and therefore to have larger black hole masses than lower power BL Lacs.  In this scenario we would expect to find a trend like the blazar sequence where powerful objects are Compton-dominant and have lower peak frequencies of emission than less powerful BL Lac type objects with smaller jet radii (due to smaller black hole masses) and correspondingly higher magnetic field strengths.  We interpret this as due to the bulk Lorentz factor of the jet, redshift and the black hole mass (which determines the size of the jet).  If we assume a jet geometry similar to that observed in M87 we expect the jet to transition from parabolic to conical at $10^{5}R_{s}$.  The jet radius at this point is a function only of black hole mass (if we assume the M87 geometry scaled linearly with black hole mass) and so the magnetic field strength and synchrotron peak frequency decrease with increasing black hole mass.  For BL Lac type objects with low black hole masses the radius of the jet at the point where it transitions to conical and is in equipartition is smaller than in a more powerful source with a higher black hole mass and this results in the increased magnetic field strength.  An empirical correlation between jet power and black hole mass has been found by several authors by comparing observed radio luminosity against black hole mass for AGN (\cite{2001ApJ...551L..17L}, \cite{2004MNRAS.353L..45M} and \cite{2006ApJ...637..669L}), however, this correlation has been disputed (\cite{2002ApJ...576...81O}, \cite{2002ApJ...579..530W} and \cite{2002ApJ...564..120H}).

If we assume that all AGN jets have approximately the same geometry as the M87 jet we can derive a simple relationship for the peak synchrotron frequency in blazars.  In fitting to the optically thick to thin synchrotron break of these blazars we are essentially fitting the size of the jet radius where the synchrotron emission is brightest.  In our model we then associate this emitting region with the region where the jet transitions from parabolic to conical.  The determination of this radius from fitting to a spectrum is a rigorous result which holds independently of the particular M87 jet geometry we have chosen.  In this work our assumption that the geometry of the jet scales linearly with black hole mass is based on the expectation that the radius of the jet close to the black hole scales with the size of the black hole event horizon which depends linearly on the black hole mass.  Of course it is likely that the geometry of individual jets depends on other quantities such as jet power, bulk Lorentz factor and the environment.  We do not yet know from observations or simulations exactly how these factors affect the jet geometry so we assume a simple scaling with black hole mass.  It is important to note that if the inner parabolic geometry of jets is not well described by this simple model this will mainly affect the inferred black hole mass of our fits and not the other physical parameters of the model.  This is the first investigation which has attempted to constrain the physical parameters of a sample of blazars using a realistic, extended model. 

The trend of decreasing peak frequency of inverse-Compton emission with jet power can also be explained using this model.  Low powered BL Lac type objects with low black hole masses have relatively high magnetic field strengths at the base of the conical section.  This means that the synchrotron emission will overwhelm CMB photons (especially since BL Lacs are observed at lower redshifts than FSRQs and have lower bulk Lorentz factors) and the inverse-Compton emission will be due primarily to scattering of relatively high frequency synchrotron photons.  This results in an inverse-Compton peak occurring at a higher frequency in BL Lacs than in Compton-dominant objects.  This is because the inverse-Compton peak frequency is proportional to the peak frequency of the seed photons $\nu_{\m{peak}} \propto \gamma^{2} \nu_{\m{seed}}$ (for $E_{\m{max}} \gg h\nu_{\m{peak}}$) and synchrotron photons have a higher peak frequency than CMB photons. 

For increasing black hole mass, the magnetic field strength decreases at the base of the conical section.  The synchrotron peak frequency decreases and this means the SSC peak frequency will also decrease.  For higher powered, high bulk Lorentz factor jets with lower magnetic field strengths Doppler-boosted CMB photons will now dominate the photon distribution within the jet at large distances and electrons will lose most of their energy scattering CMB photons instead of emitting synchrotron photons.  This produces the Compton-dominance of these objects and the low inverse-Compton peak frequency is due to the comparatively low CMB frequency.  

In this paper we have not attempted to explain the short timescale variability in Compton-dominant blazars but instead the quiescent emission.  The models we have investigated, which fit well to the quiescent emission, have electron acceleration occuring at large distances along the jet where the light-crossing time of the jet and electron cooling times are both much larger than the short flaring timescales observed (hours/minutes \cite{2007ApJ...664L..71A}).  These flares could, however, be produced by acceleration of electrons in the parabolic region of the jet where the radius of the jet is smaller, the magnetic field strength is higher and the jet is still within the BLR/dusty torus.  This means that short timescale variability is compatible with our jet model.  

In this investigation we have shown that a ballistic jet model which becomes particle dominated beyond a transition region located $>10$pc fits the multiwavelength data of the six most Compton-dominant blazars (after PKS0227) from \cite{2010ApJ...716...30A}.  We have demonstrated that using a realistic extended jet model allows us to gain constraints on the jet by reproducing radio observations.  In the next paper in the series we wish to investigate if our model is able to reproduce the spectra of BL Lac type blazars and whether our idea of a blazar sequence based on bulk Lorentz factor and black hole mass holds for a sample of BL Lacs.

\section{Conclusion}

In this paper we have used the realistic extended jet model from Paper II to fit to the simultaneous multi-wavelength spectra of six Compton-dominant blazars.  Our model includes a magnetically dominated accelerating parabolic base transitioning to a slowly decelerating conical jet with a jet geometry set by radio observations of M87 and consistent with simulations and theory.  Comparing different models of the conical section to the radio data, we find that the radio slope is well matched by both an adiabatic equipartition jet and a ballistic jet which starts in equipartition, at the transition from parabolic to conical, and becomes particle dominated further along the jet.  We use the optically thick to thin synchrotron break to constrain the radius of the jet where the jet first comes into equipartition.  We find that this requires a transition region located outside the dusty torus at large distances along the jet $>10\m{pc}$, consistent with a transition region occuring at $10^{5}R_{s}$ as in the jet of M87.  We calculate an analytic expression for the radius at which a plasma becomes optically thick to an observed frequency and use this to confirm our numerical results.  We then use this formula to calculate the expected frequency core-shift relations for the jet models under consideration. We find that for the adiabatic equipartition conical jet the core-shift relation, $x\propto \nu^{-k}$, has a value $k=0.69$, lower than suggested by observations.   

We find that the ballistic particle dominated jet fits very well to the observations across all wavelengths including radio observations whilst the adiabatic equipartition jet requires very high bulk Lorentz factors to reproduce the Compton-dominance of some of the blazars $\gamma_{\m{bulk}}>50$.  This is the first time a realistic, extended jet model has been used to fit to a sample of blazars.  We find that the fits to all these blazars require high power ($>10^{39}\m{W}$), high bulk Lorentz factor ($>20$) jets observed close to the line of sight ($<2^{o}$) consistent with the blazar sequence and unification and in agreement with the results from Paper II.  We find that the inverse-Compton emission of these objects is well fitted by scattering of CMB seed photons due to the redshift dependence of the CMB and strong Doppler-boosting.  We find that all these blazars have large inferred black hole masses ($>10^{9}M_{\odot}$) using a jet geometry set by radio observations of M87 scaled linearly with black hole mass.  For the blazar J0531 we find that the black hole mass inferred from fitting the jet model to the optically thick to thin synchrotron break agrees with the black hole mass estimated independently from the prominent accretion disc spectrum only if the transition region occurs at $10^{5}R_{s}$ as in M87.  For these large black hole masses, consistent with massive elliptical galaxies, we find that the brightest part of the jet is located at $10^{5}R_{s}$, outside the broad line region and dusty torus.

We find a relatively low value for the electron distribution power law index from fitting to the steep optically thin synchrotron emission and inverse-Compton peak ($\alpha<1.9$), implying acceleration by oblique shocks or magnetic reconnection.  We postulate a new interpretation of the blazar sequence in terms of increasing black hole mass, jet power and bulk Lorentz factor.  Higher power blazars have larger black hole masses and larger radius jets with lower magnetic fields than low power BL Lacs.  This results in high power blazars having lower synchrotron peak frequencies and less SSC emission than low power blazars with higher magnetic fields.  High power blazars also have large bulk Lorentz factors and so scatter high-redshift CMB photons at large distances resulting in the low inverse-Compton peak frequency of Compton-dominant blazars. 

These results illustrate the importance of using a realistic, physically motivated model for blazar jets to calculate their spectra and estimate their physical properties.  In a future paper we will test these concepts further by fitting to the spectra of a sample of BL Lac type objects.

\section*{Acknowledgements}

WJP acknowledges an STFC research studentship. GC acknowledges support from STFC rolling grant ST/H002456/1.  We would like to thank Roger Blandford, Jonathan McKinney and Greg Madejski for useful discussions.

\bibliographystyle{mn2e}
\bibliography{Jetpaper2refs}
\bibdata{Jetpaper2refs}

\label{lastpage}

\end{document}